\documentclass[12pt]{article}

\newcommand{\q}{\begin{equation}}
\newcommand{\qq}{\end{equation}}
\newcommand{\w}{\widehat}
\newcommand{\p}{\partial}
\renewcommand{\a}{\alpha}
\newcommand{\bb}{\beta}
\newcommand{\g}{\gamma}
\newcommand{\G}{\Gamma}
\newcommand{\dd}{\delta}
\newcommand{\m}{\mu}
\newcommand{\n}{\nu}
\newcommand{\tfrac}[2]{{\textstyle\frac{#1}{#2}}}
\renewcommand{\r}{\rho}
\newcommand{\s}{\sigma}
\newcommand{\gh}{g_{\m\n}}
\newcommand{\GG}{g^{\,\m\n}}

\newcommand{\pp}{\prime}
\newcommand{\pa}{\s_\m^{B\dot V}}
\newcommand{\h}{\tfrac12}

\renewcommand{\k}{\left |\h\,\h\right >}
\newcommand{\kk}{\left |\h,-\h\right >}
\newcommand{\am}{A_\m^{00}}
\newcommand{\oo}{\overrightarrow}
\newcommand{\wdv}{{\dot V}}
\newcommand{\rr}{\right >}

\newcommand{\llle}{\left |}
\newcommand{\la}{\lambda}

\newcommand{\gggm}{g^{\,\mu\nu}_{00}}

\begin{document}

\vspace*{2in}
\begin{center}
{\bf General Relativity Contains the Standard Model}\\[.1in]
Peter Gillan\\
gillan@alumni.fdu.edu\\
c/o Peter Schaeffer, T-BE2-03\\
Fairleigh Dickinson University, 1000 River Road,
Teaneck, New Jersey  07666, USA\\[.1in]

{\bf Abstract}

\end{center}

The Standard Model plus gravitation is derived from general relativity
with three dimensions of time. I claim that when the Lagrangian for general relativity is calculated using 
three dimensions of time, the unified field theory results. 
I call it 3D time, which stands for three dimensions of time. 
This theory differs from other higher-dimensional theories because it allows fields to
depend upon the higher-dimensional coordinates. It shows how predictions 
at the Planck mass can be tested at low energies.  The hierarchy problem is solved 
using an equation found in a classic textbook.  The theory of 3D time provides an explanation
for the masses of the electron, muon and tau, the value of the fine structure constant, 
the masses of the neutrinos of the electron, muon and tau
and the masses of the W and Z and the photon. Quark confinement 
and asymptotic freedom are produced.
The relationship between quantum mechanics and general relativity is demonstrated. 
3D time solves all these problems and many more without introducing any new problems.
The known elementary particles are solutions
to the field equations generated by the Lagrangian for general relativity.
All fields come from the same place, the metric tensor. 3D time predicts there are no supersymmetric particles.
Instead, it predicts that there are seven new intermediate vector bosons with the approximate masses
4.56 TeV, 7.32 TeV, 27.36 TeV, 29.43 TeV, 31.22 TeV, 33.04 TeV and 38.79 TeV.
Elementary tensor particles, one of which has a mass of about 58 GeV, are predicted.

\clearpage

\begin{center}

{\bf Part I. The Classical Theory}\\[.1in]

{\bf 1. Introduction}
\end{center}

Einstein added one dimension of time to the three dimensions of space.
It is only natural to add two more dimensions of time, 
bringing the number up to three.
I claim there are three dimensions of time because there are three dimensions of space. 
That there are three dimensions of time is the first postulate upon which this work is based.
This will finish the job Einstein started of adding dimensions of time and 
making time symmetrical to space.

To specify a theory of three dimensions of time (3D time), one need look no
further than the general theory of relativity.  The Lagrangian for any theory contains 
all of that theory's physical information. 
The Lagrangian for general relativity is the curvature scalar. 
The Lagrangian for 3D time is also the curvature scalar. 
This is the second and last postulate upon which this work is based.

The theory of 3D time is general relativity because it has the curvature scalar as its Lagrangian.
This means it is the most beautiful theory ever constructed because no theory is more
beautiful than general relativity and that is what 3D time is. And it is relativity in another way.
It is a theory of pure gravitation because its Lagrangian is the curvature scalar alone.
The theory of 3D time continues the over-300-year-old tradition started by Newton 
and continued by Einstein of placing gravitation in the position of greatest importance. 
History repeats itself.

The postulate of three dimensions of time implies a six-dimensional spacetime.
The six-dimensional curvature scalar $\w R$ will give a complete description 
of the elementary particles. One starts off calculating the Christoffel symbols 
$\w \Gamma^\alpha_{\beta\gamma}$ and winds up describing the elementary particles.
This document is at the same time both general relativity and high energy physics,
a necessary condition for a unified field theory.

With widespread uncertainty about the current state of theoretical physics,
one may wonder whether strings and supersymmetry are correct. Indeed,
everything that follows from a faulty initial assumption is almost certainly wrong.
One may wonder if there are better initial assumptions. I claim there are no better 
ones than gravitation and three dimensions of time. With the vision of 
Albert Einstein and Isaac Newton backing them up, they are prime candidates for a 
unified field theory. Great minds think alike.

Our story begins with the Kaluza-Klein (KK) [1,2] miracle. 
If one assumes fields do not depend on the higher-dimensional coordinates,
then the curvature scalar in five dimensions reduces to the
curvature scalar in four dimensions plus the term for Maxwell's equations.
It seems to be a miracle that gravitation in five dimensions yields 
gravitation in four dimensions plus electromagnetism. However,
all other fields and terms that we now know
exist do not appear in the Lagrangian.  It turns out that if the artificial
assumption that fields do not depend on the higher dimensional coordinates
is eliminated, the fields and terms of the Standard Model (SM) [3-6] appear.
Indeed fields should depend on all coordinates, including the higher-dimensional ones,
so making the theory more self-consistent also makes it apply to all particles and fields.
This is the best of both worlds. It works, and the only thing left to explain is why this has not been done before.
The answer is that along with the terms of the SM, hundreds of other weird looking terms appear, obscuring
the physically important ones. Most of these weird terms are negligible as I explain later.

Klein postulated that the fifth dimension was microscopic and closed,
resulting in an infinite tower of harmonics. This tower constituted a field's
higher-dimensional (HD) coordinate dependence.
The fundamental was massless.  The higher order particles, which
represented the dependence on the fifth coordinate, had Planck masses.
In 3D time, this dependence will produce the fields and terms that
we now know exist. This will be the realization of
Einstein's dream [7--9] of accounting for all physical phenomena
from the ``pure marble" of geometry, without the ``base wood"
of additional matter fields. Many of the terms in the Lagrangian for the
SM each represent a separate assumption. These terms will be produced by 
just one assumption, the curvature scalar $\w R$ of 3D time.  
The SM is a Lagrangian that can be derived from the curvature scalar.

Previously, it was thought that fields could not depend on the HD coordinates 
because they would be too massive to detect.
However, 3D time solves the hierarchy problem so the 
Planck masses of these massive harmonics are reduced to ordinary 
elementary particle masses.  This helps produce the familiar elementary particles.
See Table~1.
 
\begin{center}
\begin{tabular}{ccccccccccccccc} 
         
         \hline\hline
        	$0$& & & & & & & &$\gamma$& & & & & & \\ 
		& & & & & & & & & & & & & & \\
		$\tfrac12$& & & & & & &$e^-$& &\ $\nu_e$& & & & & \\
		& & & & & & & & & & & & & & \\
		$1$& & & & & &$W^-$& &$Z^0$& &\ $W^+$& & & & \\
		& & & & & & & & & & & & & & \\
	      $\tfrac32$& & & & &$\nu_{2\mu}$& &$\mu^-$& &\ $\nu_{\mu}$& 
                          &\ $\mu_2^+$& & & \\
		& & & & & & & & & & & & & & \\
		$2$& & & &$A_\mu^{2,-2}$& &$A_\mu^{2,-1}$& &$A_\mu^{20}$& 
                      &\ $A_\mu^{21}$& &\ $A_\mu^{22}$& & \\
		& & & & & & & & & & & & & & \\
		$\tfrac52$& & &$\nu_{3\tau}$& &$\nu_{2\tau}$&
                                 &$\tau^-$& &\ $\nu_{\tau}$&
                                 &\ $\tau_2^+$& &\ $\nu_{4\tau}$& \\
		& & & & & & & & & & & & & & \\ 
		$3$& &$A_\mu^{3,-3}$& &$A_\mu^{3,-2}$& 
                &$A_\mu^{3,-1}$& &$A_\mu^{30}$& &\ $A_\mu^{31}$& 
                &\ $A_\mu^{32}$& &\ $A_\mu^{33}$\\
		& & & & & & & & & & & & & & \\  [.03in] \hline
		$l$&$/ m$&$-3$&$-\tfrac52$&$-2$&$-\tfrac32$&
          	$-1$&$-\tfrac12$&$0$&$\ \tfrac12$&$\ 1$&
          	$\ \tfrac32$&$\ 2$&$\ \tfrac52$&$\ 3$\\
  \hline\hline
\end{tabular}
\end{center}
Table~1. Tower of harmonics in 3D time.
The vectors and leptons have quantum numbers $|lm\rangle$.
\vspace{0.1in}

This conversion of Planck masses to elementary particle masses allows predictions at the 
Planck mass to be tested at low energies. The conversion is allowed by
quantum field theory that is simplified by particles that are not points, but are
line segments of charge. The line segments spin, producing a three-dimensional
distribution of charge. Presumably, these particles 
avoid the violation of causality or special relativity usually associated with extended particles 
just like the particles of string theory, but this is conjecture.

Nine masses and one coupling constant are derived in Part~II.
Neither the symmetry group $SU(3)\times SU(2)\times U(1)$ nor the 
Higgs mechanism will be derived. These are shown to be superfluous in Part~II.
They are not needed because particles in 3D time are not points.
Point particles require the condition of renormalizability, which is preserved by
gauge symmetry and the Higgs mechanism. 
If particles are extended, as claimed by 3D time and string theory, renormalizability
and the gauge symmetry that preserve it become unnecessary.
Likewise, the Higgs mechanism is not needed.
This is unlike today's Kaluza-Klein theories, where the observed elementary
particles lie in the zero-mass sector of a Kaluza-Klein theory and receive
a mass from the Higgs mechanism. In this case a determination of the masses of the
elementary particles is out of the question.  They are completely unexplained.
In 3D time, however, complete explanations of masses are given.

The organization of Part I is as follows:   
Section~2 presents a comparison of 3D time with other Kaluza-Klein theories.
Section~3 breaks down the six-dimensional (6D) metric tensor into
four-dimensional and HD quantities. Section~4 specifies the Lagrangian.
Four-dimensional gravitation is derived in Sec.\,5, where a discussion
on elementary tensor particles can be found.  
In Sec.\,6, the term for Maxwell's equations and mass terms
for the photon, W and Z are obtained. Methods for dealing with
the HD coordinates in each term in the Lagrangian
are given in Sec.\,7. Section~8 introduces fermions and derives the
Dirac equation from the 6D curvature scalar.
Terms for leptons interacting with the W, Z and photon are deduced
in Sec.\,9. Section~10 derives the strong interactions.
Quarks, confinement, asymptotic freedom and chiral symmetry
breaking are produced in Sec.\,11.
\pagebreak[2]

{\samepage
\begin{center}
{\bf  2. Kaluza-Klein theories}
\end{center}

\nopagebreak
I will now present a comparison of 3D time with other Kaluza-Klein (KK)  
theories.  This will follow closely the review by Overduin and Wesson [10].
The single biggest difference between 3D time and other KK theories
is that fields in 3D time depend on the higher-dimensional coordinates.}

DeWitt [11] was the first to suggest incorporating the
non-Abelian symmetry $SU(2)$ of the SM into a KK theory.
Others [12--14] took up the challenge, ending with Cho and Freund [15,16].
The main difference between these KK theories
and 3D time is the origin of vectors in the theories.
In the former, each dimension produces one vector in the usual way.
In 3D time, one dimension produces all vectors needed for the SM.
This is made possible by the expansion of the lone vector in terms of spherical harmonics.
This represents the vector's dependence upon the HD coordinates $\theta$ and $\phi$, 
the familiar spherical coordinates.
This procedure is superior because it accounts for the
$SU(2)\times U(1)$ structure of electroweak symmetry.
Here the photon is associated with the lowest order $l=m=0$
spherical harmonic and the $W^\pm$ and $Z^0$ are connected
with the next higher order $l=1$, $m=-1,0,1$ harmonics.
See Table~1. 

The disadvantage of previous non-Abelian KK theories is that they
naively require one dimension to produce each vector. Thus,
many higher dimensions are required to produce the group
$SU(3)\times SU(2)\times U(1)$ of the SM. In addition, there is no
reason why some of the higher dimensions should be different than others.
This would be required to explain the group
$SU(3)\times SU(2)\times U(1)$. In particular, the question
of why the photon is different than the W and Z 
is left unanswered. Likewise for the difference between the 
strong and electroweak interactions.

The theory of 3D time and these non-Abelian KK theories, however, are similar in the
following respect: All matter fields are contained within the
6D curvature scalar. There are no external, additional
matter fields. This embodies Einstein's vision of
nature being the result of pure geometry. Likewise, each of these
KK theories is pure gravitation with indices allowed to
run to values greater than four.

References [17--24] give the physical implications of HD coordinate dependence for
general relativity in the macroscopic realm --- such as its effect on the
advance of the perihelion of Mercury. The theory of 3D time
gives its implications for microscopic physics  --- elementary particles.

Compactification in non-Abelian KK theories can be a problem. However,
with only two higher dimensions that form an ordinary sphere,
compactification in 3D time is trivial, just as it is for the original
KK theory with one higher dimension. We do not have to worry about
compactification as in, for example, the theories of Cho and Freund
[25--29], in which the higher dimensions have general curvature.
In this case, one runs into difficulties with the consistency
of the field equations when the dependence upon the HD coordinates is
eliminated (Duff [30--33]). The theory of 3D time does not have this
problem. Even if it did we could solve it by allowing fields to depend on the
HD coordinates and agree with Cho's call [34--37]
for abandonment of the ``zero modes approximation."

In 3D time, one does not need any of the various compactification mechanisms
such as altering Einstein's equations by incorporating torsion [38--41],
adding higher-derivative terms such as $R^2$ to the Lagrangian [42],
or by adding matter fields.  This last method, with the right matter
terms, is known as spontaneous compactification.  The first example of 
this is that of Cremmer and Scherk [43,44].

All previous unified field theories that include gravity (theories of everything)
suffer from the same problem --- the inability to make sufficient
contact with low-energy phenomenology.
In other words, the theories are not realistic. These include
the original KK theory in five dimensions, the extensions
of this theory with more than one higher dimension (intended to account
for non-Abelian symmetry), Supergravity, string theory,
and the present-day theory of everything, M-theory.

For example, none of these theories produces the right particle spectrum.
The KK theories without supersymmetry have no fermions.
The incorporation of fermions in such theories is apparently without explanation.
Supersymmetry meliorates the situation. However in this case,
the most abundant type of particles is scalars.  The theory of d=11, N=8 Supergravity
has 70 of them.  This is in contradiction to observation, which indicates
that there are few scalars.  The presently-accepted way out is to assume
that the scalars are too massive to be detected.  This means 70
additional assumptions. This is unsightly, to say the least.

The solution to this problem is to incorporate the branch of mathematics
known as spinor theory [45] into KK theory.  Spinor theory dictates
that a null vector (the photon) may be equated to a pair of spinors
(fermions).  When this is combined with 6D KK theory with the proper
HD coordinate dependence, all of the observed fermions
are produced. And because there is no relation between fermions and
scalars as with supersymmetry, this produces no scalars. 
Agreement with observation is at hand.

The current method of choice for the incorporation of fermions into
KK theory is supersymmetry.  This is more natural than their 
incorporation by hand, but there are severe drawbacks.
For example, the fermionic superpartners of KK's bosons have not been 
detected.  Similarly, the bosonic superpartners of the observed fermions
are not observed.  Thus there is little evidence from experiment
for supersymmetry.  By comparison, the evidence for
3D time is the fermionic physics of the SM --- it will reproduce
the observable features of the SM.  Spinor theory
dictates that the photon may be equated to many pairs of
fermions in the Lagrangian. These fermions are the observed ones.
Spinor theory in 3D time is proven
every day in particle accelerators. It will account for pair production 
by producing spinor-vector interaction terms in the Lagrangian
(the assumption of minimal coupling in the SM).  With two spinors and one vector
as factors, this term represents a Feynman diagram with two
external fermionic lines and one external bosonic line.

In addition, one of the new terms to appear naturally in the 6D curvature scalar as a result
of HD coordinate dependence is the term for the Dirac equation.
This equation and all it implies must be assumed in other theories
such as $d=11$ Supergravity, for example.

One advantage of Supergravity is that it provides a number for the
dimensionality of spacetime.  Nahm [46] showed 11 is the 
maximum number of dimensions of spacetime for supersymmetric gravity.
In 11 dimensions, nature is maximally supersymmetric.
Witten [47] proved 11 is the minimum number of dimensions
required for the incorporation of the group $SU(3)\times SU(2)\times U(1)$
of the SM.  However, this assumes the one-vector-for-one-dimension
aspect, along with its disadvantages, of the non-Abelian theories mentioned above.
Cremmer, Julia and Scherk [48] showed that in 11 dimensions
there is only one choice for additional matter fields.
Freund and Rubin [49] showed that d=11 Supergravity naturally compactifies
to four macroscopic and seven microscopic dimensions.

3D time also provides a number for the dimensionality of spacetime.
Instead of a symmetry between bosons and fermions, it postulates a
symmetry between space and time. This implies there are
three dimensions of time. There are three dimensions
of time because there are three dimensions of space.  Spacetime has
six dimensions.  In six dimensions all the observable features of the Standard Model, 
in all their complexity and with all their idiosyncracies, will be produced.
We choose six dimensions because it reproduces known physics.

The theory of $d=11$ Supergravity lost its status as the theory of everything
in the mid 1980's.
There were several reasons for this.  First, the extra spacetime
did not contain quarks or leptons, nor the gauge group for the SM.
Second, one cannot build chirality, necessary for an accurate description
of fermions in the Standard Model, into a theory with an odd number of
dimensions.  Third, it had a large cosmological constant, contradicting
observation. Fourth, it had anomalies.
3D time naturally accounts for what appears to be the gauge group
for the SM with its expansions in terms of spherical harmonics
for electroweak theory and the HD part of the 6D rotation group for
the strong interactions.  In 3D time, these expansions, together
with spinor theory, do indeed produce quarks and leptons.
In six dimensions, chirality is possible.  The large cosmological
constant appearing in the 4D version of Kaluza-Klein theory
is naturally eliminated in 3D time. Finally, it will be easy to show
3D time is anomaly-free.

String theory [50] is supposed to be able to predict the parameters of the
Standard Model.  It does not do this. I will show 3D time does this.
There are about 20 unexplained parameters in the SM. 3D time
explains ten of these at the present time.  String theory
explains none.

String theory has three separate uniqueness problems.
First, the number of supersymmetric generators can have any value
from N=1 to N=8. Second, there are five different classes of string
theory. Finally, within each class there are thousands of different
string theories, each corresponding to a different way one can
compactify seven higher dimensions. In contrast, there is only one 
theory of 3D time. 3D time does not have supersymmetry,
with its multiple values for N.  Particles in 3D time are not strings,
which would be classified into five types. Finally, there is only
one way to compactify the two higher dimensions.  It is a trivial compactification
to the 2-sphere, similar to the compactification of the one HD of original
Kaluza-Klein to the 1-sphere or circle.  Therefore, 3D time does not suffer
from uniqueness problems.

String theory has matter terms
known as Chapline-Manton terms [51] that must be added to the
Lagrangian.  In six dimensions, 3D time has no matter terms.

Green and Schwarz [52] and Gross et al. [53] showed all anomalies
vanish for SO(32) and $E_8\times E_8$ string theory, respectively.
String theory may provide an anomaly-free path to quantum gravity [54].
However, I claim quantum gravity is not the unification of gravity and quantum theory.
I can produce a unification of the Standard Model and gravitation without it.
How can a theory whose effects are entirely negligible be important?
The relationship between gravity and quantum theory is given
below. But if you must have a quantum theory of gravity,
my new way of doing quantum field theory will help.

This being said,
I will now show why 3D time is anomaly-free.
First, as will be shown below, the 6D metric tensor
in this theory contains all types of 4D matter fields.
In addition, the 6D curvature scalar contains all types of 4D
matter terms.  These are accomplished by expanding certain components of the metric tensor
in terms of spherical harmonics, spinors and 6D
rotation group generators. Now, the Lagrangian for 3D time
is the 6D curvature scalar.
There are no other terms, which could be called 6D matter terms.
In other words, in six dimensions, spacetime is empty. There are no
currents. Thus, there are no conservation laws.
Now, anomalies are quantum mechanical violations
of conservation laws.
Without conservation laws there are no anomalies.
Therefore, 3D time is anomaly-free.

The worst disadvantage of string theory is that it does not make
predictions. They reside at the Planck mass.
M-theory is worse than string theory because it is nonperturbative as well.
The electroweak sector, which contains the testable predictions
of 3D time, is perturbative. More importantly, 3D time has a mechanism
for converting Planck masses to elementary particle masses.
It solves the hierarchy problem.
Thus, it makes clear-cut physical predictions.  3D time predicts the
observable features of the Standard Model as will be shown. In addition,
it predicts new intermediate vector bosons with precisely determined masses
as well as several generations of scalars, the first one of which has been detected.
None of these scalars is the Higgs scalar, however. 3D time has a much better method
of generating masses than the Higgs mechanism. The method produces the masses
of nine elementary particles. The results speak for themselves.
\pagebreak[2]

{\samepage
\begin{center}
{\bf  3. The 6D metric tensor}
\end{center}

\nopagebreak
In this section, we break down the 6D metric tensor into 4D and HD
quantities in preparation for its substitution into the 6D
curvature scalar.  The postulate of three dimensions of time implies a
6D spacetime.  Included in this postulate of six dimensions
is the size, shape, spacelike or timelike nature and
connection to 4D spacetime of the higher dimensions.  Therefore, 
we will further postulate that the two HD's form the surface of a sphere
with radius $T$. This radius is the
Planck time $T$.  We have
\begin{eqnarray}
T&=&5.39\times10^{-44}\ \mbox{sec},\label{eq:2.11z}\\
cT&=&L=1.62\times 10^{-33}\ \mbox{cm},\label{eq:15.6}
\end{eqnarray}
where $L$ is the Planck length and $c$ is the speed of light.}
I claim the Planck length is important because the radius of the
higher-dimensional sphere is the Planck time. 
The reason the two higher dimensions are so small is cosmological
in origin. Apparently, while the dimensions of 4D spacetime have been
expanding, the two higher dimensions have been shrinking.

Hawking [55] proposed the use of imaginary time. That is, one uses the coordinate $it$
instead of $t$, where $i=\sqrt{-1}$. This is only natural, because one can not see time,
unlike space, so it exists only in one's imagination. Therefore, it is imaginary. 
One is allowed to multiply the coordinate $t$ by $i$
because multiplying a coordinate by a constant amounts to choosing one's coordinate system,
which one is free to do. The following example illustrates this.

Consider a two-dimensional spacetime with coordinates $x$ and $t$. We have the line element
$$ds^2 = g_{ij}dx^{\,i} dx^{\,j} = dxdx-c^2 dtdt.\label{eq:2.1}$$
If instead one uses the coordinates $x^{\,1} = x$ and $x^{\,2} = ict$ for the same spacetime,
one arrives at the line element
$$ds^2 = g_{ij}dx^{\,i} dx^{\,j} = dx^{\,1} dx^{\,1} + dx^{\,2} dx^{\,2}.\label{eq:2.1a}$$
As 't Hooft [56] noted, the coordinate $t$ is timelike because $g_{22}$ for this coordinate is negative,
while the coordinate $x^{\,2} = ict$ is spacelike because $g_{22}$ for this coordinate is positive.
Thus, we have incorporated the minus sign usually associated with timelike diagonal components
of the metric tensor into the coordinate of time, resulting in a positive sign for these components.

It is known that timelike higher dimensions lead to a negative-energy electromagnetic field.
In addition, 't Hooft [56] related that timelike higher dimensions lead 
to unbounded and tachyonic masses for the elementary particles. Clearly, one must have spacelike 
higher dimensions to avoid these difficulties. So we multiply the HD coordinates of time by $i$, 
rendering the higher dimensions spacelike. 
We multiply all three coordinates of time in 3D time by $i=\sqrt{-1}$.
Therefore, like in the above example, the signature of the metric tensor is diag.\,(~+~+~+~+~+~+~). 
This simple coordinate transformation, which one is free to make,
debunks the myth that the higher dimensions must be
those of space, one of the biggest stumbling blocks to the theory of three dimensions of time.
Hawking's idea of imaginary time may solve other important problems.
It may eliminate the singularity at the center of a black hole and at the beginning of the universe.

The differential line element for the two coordinates of time $\theta$ and $\phi$ that 
form a sphere of radius $T$ is
\begin{equation}
ds^2=\w g_{ij}d\w x^{\,i}d\w x^{\,j}=
     -c^2 T^2\,d\theta^2-c^2 T^2\sin^2\!\!\theta\,d\phi^2,\label{eq:2.2}
\end{equation}
where $i,j$ take the value five or six and $\theta$ and $\phi$ are
the usual spherical coordinates. A hat over a symbol denotes a 6D quantity.
However, instead of choosing $\w x^{\,5}=\theta$ and $\w x^{\,6}=\phi$ with
$\w g_{55}=-c^2T^2$ and $\w g_{66}=-c^2T^2\sin^2\!\!\theta$,
we set
\begin{eqnarray}
\widehat x^{\,5}&=&icT\,\theta,\label{eq:2.2a}\\
\widehat x^{\,6}&=&icT(\sin\!\theta)\,\phi,\label{eq:2.2b}\\
\widehat g_{55}&=&\widehat g_{66}=1,\label{eq:2.2c}\\
\widehat g_{56}&=&\widehat g_{65}=0.\label{eq:2.2d}
\end{eqnarray}
These assignments for the HD coordinates lead to the same line element Eq.\,(\ref{eq:2.2})
and follow the definition of the metric tensor as the dot product of unit vectors.

The incorporation of the factors of
$cT$ into $\w x^{\,5}$ and $\w x^{\,6}$
makes the HD coordinates more like the
4D ones, which have the units of distance. This suggests
we incorporate the factor of $\sin\!\theta$ into $\w x^{\,6}$ instead
of $\w g_{66}$. This eliminates a cosmological constant
while keeping the higher dimensions small. This is described in Sec.\,4.
The value 1 for $\widehat g_{66}$ gives the appearance of what is called
Ricci-flatness of the HD's.
This means [31] the cosmological constant is zero and,
with the elimination of the scalar field $\widehat g_{66}$, there is no
Higgs mechanism. Both of these features are not typical of today's
KK theories.

For use in calculating the Lagrangian in Sec.\,4, We tabulate here the
differentials and derivatives of the two HD
coordinates $\w x^{\,5}$ and $\w x^{\,6}$ according to
Eqs.\,(\ref{eq:2.2})--(\ref{eq:2.2b}) 
\begin{eqnarray}
d\w x^{\,5}&=&icT\,d\theta,\label{eq:2.3a}\\ 
\p_5=\frac{\partial}{\partial\w x^{\,5}}&=&
\frac{1}{icT}\frac{\partial}{\partial\theta},\label{eq:2.3b}\\
d\w x^{\,6}&=&icT\sin\!\theta\,d\phi,\label{eq:2.3c}\\
\p_6=\frac{\partial}{\partial\widehat x^{\,6}}
&=&\frac{1}{icT\sin\!\theta}\frac{\partial}{\partial\phi}.\label{eq:2.3d}
\end{eqnarray}

The 6D metric tensor is denoted by
$\w g_{\alpha\beta}$,
where indices in the
beginning of the Greek alphabet such as $\alpha,\ \beta,\ \gamma,$
and $\delta$ run from one
to six.  The indices $\mu,\ \nu,\ \rho,$ 
and $\sigma$ in the middle of the Greek alphabet range from
one to four. The first four coordinates
of the 6D spacetime are those of the ordinary 4D
spacetime of experience.

We postulate the connection between the HD sphere and 4D spacetime as follows:
The 4D worldline of the particle under consideration is the $z$-axis of the 3D
embedding space of the HD sphere. This means the $z$-axis of the embedding space of
the HD sphere is a line of 4D spacetime.
This overlap of the 4D and HD spaces is necessary because if one assumes
an infinitesimal surface area perpendicular to a line in 4D spacetime is part of the
HD sphere, one is inescapably lead to the conclusion that the surface must curve
around the line. The line and the sphere then exist in the same 3D space.

The circular sixth dimension, parameterized by the coordinate $\phi$,
is perpendicular to the $z$-axis.
Taking this axis to be $x^1, x^2, x^3$, or $x^4$, we have $\w g_{6\mu}=0$.
The coordinate $\phi$, along with a line of 4D
spacetime, forms a cylinder, and the metric tensor for these perpendicular coordinates
is zero. It is not realized that this most common type of HD geometry
causes the vector $\w g_{5\mu}$ in the original KK theory
to be zero. The fifth dimension in 3D time is parameterized by the coordinate $\theta$.
It is measured from the positive $z$-axis. It is not,
in general, perpendicular to the $z$-direction consisting of 4D spacetime.
Therefore, the vector $\w g_{5\mu}$ is not zero. This is responsible for the 
existence of nearly all elementary particles, which will be contained in the vector
$A_\mu=\w g_{5\mu}$. It is only because of this peculiar overlap of the 4D and HD spaces
that $A_\mu$ and all matter is not zero.

Allowing fields to depend on the HD coordinates will produce the fields and terms 
necessary for the Standard Model. At the same time, this makes 3D time more self-consistent
and correct because an arbitrary field should depend on all six coordinates, not just the first four. 
The two HD coordinates should be treated like the first four coordinates, upon which fields depend.
This removes the ad-hoc assumption in Kaluza-Klein theory that
fields do not depend on the HD coordinates. This dependence upon the HD coordinates
is represented by a Fourier expansion in terms of
spherical harmonics.  We have
\begin{equation}
g_{\mu\nu}(\w x^{\,\alpha})=
\sum_{l=0}^\infty
\sum_{m=-l}^l
g_{\mu\nu}^{lm}(x^\rho)
Y_{lm}(\theta,\phi),\label{eq:2.9a}
\end{equation}
\begin{equation}
A_\mu(\w x^{\,\alpha})=
\sum_{l=0}^\infty
\sum_{m=-l}^l
A_\mu^{lm}(x^\nu)
Y_{lm}(\theta,\phi),\label{eq:2.9b}
\end{equation}
where the $Y_{lm}(\theta,\phi)$ are the spherical harmonics.

The 6D metric tensor is defined in terms of 4D quantities as follows
\begin{equation}
\left(
\begin{array}{ccc}
\w g_{\mu\nu}&\w g_{\mu5}&\w g_{\mu6}\\
\w g_{5\nu}&\w g_{55}&\w g_{56}\\
\w g_{6\nu}&\w g_{65}&\w g_{66}
\end{array}
\right)=\left(
\begin{array}{ccc}
g_{\mu\nu}+A_\mu A_\nu&A_\mu&0\\
A_\nu&1&0\\
0&0&1
\end{array}
\right).\label{eq:2.8a}
\end{equation}
For an example of how to obtain an equation like (\ref{eq:2.8a}),
see Ref.\,[57].

The contravariant metric tensor, introduced via the relations
$\w g^{\,\alpha\beta}\w g_{\beta\gamma}=
\w \delta_{\,\ \gamma}^{\,\alpha},$
is
\begin{equation}
\left(
\begin{array}{ccc}
\w g^{\,\mu\nu}&\w g^{\,\mu5}&\w g^{\,\mu6}\\
\w g^{\,5\nu}&\w g^{\,55}&\w g^{\,56}\\
\w g^{\,6\nu}&\w g^{\,65}&\w g^{\,66}
\end{array}
\right)=\left(
\begin{array}{ccc}
g^{\mu\nu}&-A^\mu&0\\
-A^\nu&
1+A^\mu A_\mu&0\\
0&0&1
\end{array}
\right).\label{eq:2.8b}
\end{equation}
\pagebreak[2]

{\samepage
\begin{center}
{\bf 4. The Lagrangian}
\end{center}

\nopagebreak
The Lagrangian density for 3D time is 
\begin{equation}
{\cal L}=k\w R,\label{eq:3.1}
\end{equation}
where $\w R$
is the 6D curvature scalar and 
$k=c^3/16\pi G$, where $c$ is the speed of light and $G$ is the
constant of gravitation.}
We have
\begin{equation}
\w R=\w g^{\,\a\bb}\w R_{\a\bb},\label{eq:3.1a}
\end{equation}
where $\w R_{\a\bb}$ is defined in the usual way
\begin{equation}
\w R_{\a\bb}=
\p_\g\w\G^{\,\g}_{\,\ \a\bb}
-\p_\bb\w\G^{\,\g}_{\,\ \a\g}
-\w\G^{\,\dd}_{\ \,\bb\g}\w\G^{\,\g}_{\ \,\a\dd}
+\w\G^{\,\dd}_{\,\ \a\bb}\w\G^{\,\g}_{\ \,\dd\g}.\label{eq:3.1b}
\end{equation}
Here
\begin{equation}
\w\G^{\,\g}_{\,\ \a\bb}=
\w g^{\,\g\dd}\w\G_{\dd\a\bb},\label{eq:3.1c}
\end{equation}
where

\begin{equation}
\w\G_{\dd\a\bb}=
\tfrac{1}{2}\left(
-\p_\dd\w g_{\a\bb}
+\p_\a\w g_{\bb\dd}
+\p_\bb\w g_{\dd\a}\right).
\label{eq:3.1d}
\end{equation}

The action integral for 3D time is
\begin{equation}
\w I=k\int\w R\sqrt{\w g}\,\,d^{\,6}\w x,\label{eq:3.2}
\end{equation}
where $\w g$ is the determinant of the 6D
metric tensor.  This determinant may be expressed in terms of
the 4D metric tensor $g_{\m\n}$.
If we multiply the middle column of Eq.\,(\ref{eq:2.8a})
by $A_\n$ and subtract the resulting column from the first, we obtain
\begin{equation}
\sqrt{\w g}=
\left|
\begin{array}{ccc}
g_{\m\n}&A_\m&0\\
0&1&0\\
0&0&1
\end{array}\right|^{1/2}=
\sqrt{g}\,.
\label{eq:3.2a}
\end{equation}
Using Eqs.\,(\ref{eq:3.2a}),~(\ref{eq:2.3a}) and~(\ref{eq:2.3c}), 
the 6D volume element becomes
$-\sqrt{g}\,c^2 T^2\sin\!\theta\,d\theta\,d\phi\,d^4x$.
The constant $-c^2 T^2$ may be ignored
as the entire Lagrangian is multiplied by it and the field
equations remain the same after its removal.

The dependence upon the two HD coordinates has a profound effect
on 6D Kaluza-Klein theory.  Instead of obtaining only the terms for
4D gravitation and electromagnetism (the Kaluza-Klein miracle),
one now obtains literally hundreds of new, odd-looking terms.
This is because terms with $\partial_5$ or $\partial_6$ are not zero now.
This embarrassment of riches is apparently why Kaluza did not
allow dependence upon the extra coordinate. Certainly, it is why
the full-fledged theory was rejected. Why bother to look for
physically important terms when to all outward appearances you have
done something wrong?

The shortcut to understanding what these terms mean
is to consider those with as few factors as possible.
The proper procedure starts by determining the
free-field terms for each field of interest.
These are obtained by setting to zero all the other fields
in each term in the Lagrangian.
Thus, the free-field terms have only
the one type of field under consideration in them.
They exclude terms with other fields in them.
These would be interaction terms, which have more than one type 
of field.

There are three classes of terms in the Lagrangian.
First are the terms for the
free-$g_{\m\n}$ field, which will contain the graviton field.
These contain only $g_{\m\n}$ or
$g^{\,\m\n}$; they are obtained by setting $A_\m=0$.  These
terms are important macroscopically, where the sources of
$g_{\m\n}$, e.g., masses, add and those for the $A_\m$ field cancel.
Second are the terms for the free-$A_\m$ field, which will contain
the photon field.  These are
obtained by setting $g_{\m\n}=\delta_{\m\n}$, which is the 
Kronecker delta.  Note one cannot set $g_{\m\n}=0$; 
however it can be frozen out by setting it equal to a constant.
This causes all derivatives of $g_{\m\n}$
and $g^{\,\m\n}$ to vanish.  These terms are important 
microscopically, where the sources of the $A_\m$ field, e.g., charges, do not
cancel and those for the $g_{\m\n}$ field are very
small.  Third are the terms for the interaction of the
$g_{\m\n}$ and $A_\m$ fields.  These contain at least one
factor of $A$ and at least one derivative of $g_{\m\n}$
or $g^{\,\m\n}$.  These terms vanish both macroscopically
and microscopically, where $A_\m$ vanishes and $g_{\m\n}$
is constant, respectively.  We will not consider this
class of terms further.
\pagebreak[2]

{\samepage
\begin{center}
{\bf 5. Four-dimensional gravitation and elementary tensor particles}
\end{center}

\nopagebreak
There are three types of free-field terms in the Lagrangian for
$g_{\mu\nu}$: the 4D curvature scalar $R$, terms with HD
derivatives with two factors of $\gh$, and terms with HD derivatives and four factors of
$\gh$. Terms with HD derivatives and
four factors of $\gh$ are self-interaction terms for the $\w g_{\m\n}^{lm}$.
Aside from these, we have
\begin{eqnarray}
&k&R\label{eq:4.1}\\
-&k&\GG\p_5^{\ 2}\gh\label{eq:4.2}\\
-&k&\GG\p_6^{\ 2}\gh\label{eq:4.3}\\
-\tfrac{1}{2}&k&\p_5\GG\p_5\gh\label{eq:4.4}\\
-\tfrac{1}{2}&k&\p_6\GG\p_6\gh\label{eq:4.5}
\end{eqnarray}}

Because the size of the HD sphere is of the order
of the Planck length, one cannot observe the HD
coordinate dependence of $\gh$.  To eliminate it, one must
integrate over the HD coordinates, thereby
averaging over them and obtaining a 4D description of the
6D curvature scalar.

Term~(\ref{eq:4.1}) contains the set of terms for
4D gravitation.  These are obtained by setting
$\gh=g_{\m\n}^{00}|00\rangle$ and
$\GG=\overline g^{\,\m\n}_{00}\langle00|$.  One may group all the kets
together and all the bras together to yield a single bra
$\langle00|$ and ket $|00\rangle$, which when multiplied
together yield one.  The result is the curvature scalar
$R(g_{\mu\nu}^{00})$, which contains the 4D quantities $g_{\m\n}^{00}$,
$g_{00}^{\,\m\n}$, and $\partial_\mu$ only.  Therefore, we may interpret $g_{\m\n}^{00}$
as the graviton field.

The procedures for taking the derivatives with respect to
the HD coordinates and integrating over $\theta$ and $\phi$ for 
Terms~(\ref{eq:4.2})--(\ref{eq:4.5})
are the same as those for Terms~(\ref{eq:5.2})--(\ref{eq:5.4a})
described in Sec.\,7.  Considering only the first four terms in the expansion of
$g_{\mu\nu}$ in terms of spherical harmonics, we arrive at the following result
\begin{eqnarray}
+\frac{9-3\sqrt3}{12c^2 T^2}&k&
\overline g_{00}^{\,\m\r}\overline g_{00}^{\,\n\s}
g_{\r\s}^{1,-1}g_{\m\n}^{11}\label{eq:4.6a}\\
+\ \ \frac{9-3\sqrt3}{12c^2 T^2}&k&
\overline g_{00}^{\,\m\r}\overline g_{00}^{\,\n\s}
g_{\r\s}^{11}g_{\m\n}^{1,-1}\label{eq:4.6b}\\
+\ \ \frac{2\sqrt3\,+\sqrt{30}\,}{12c^2 T^2}&k&
\overline g_{00}^{\,\m\r}\overline g_{00}^{\,\n\s}
g_{\r\s}^{10}g_{\m\n}^{10},\label{eq:4.6c}
\end{eqnarray}
where we have considered only the $g_{00}^{\,\m\n}$ term in
$\GG$ and only the $g_{\m\n}^{00}$ field in the factor of
$\sqrt{g}\,$ in the differential volume element.  

As $g^{\,\m\n}_{00}$ is the graviton field, 
Terms~(\ref{eq:4.6a})--(\ref{eq:4.6c}) are mass
terms for the $g_{\m\n}^{1m}$ fields.  The field $g_{\m\n}^{00}$
does not have a mass term because its associated spherical harmonic
$Y_{00}=1/\sqrt{4\pi}$ does not depend on the HD coordinates.
Therefore, its derivative with respect to these coordinates is zero and
$g_{\m\n}^{00}$ is massless. This is consistent with our
interpretation of it as the graviton field.
Therefore the theory of 3D time contains 4D gravitation.

The tensor $g_{\m\n}^{00}$ is the 4D metric tensor
because it contains no reference to 6D spacetime. It is
completely four dimensional.
Thus, $g_{00}^{\,\m\n}$ and $g_{\m\n}^{00}$ raise and lower
4D indices, respectively. In addition, only
$g_{\m\n}^{00}$ appears in the factor of $\sqrt{g}\,$
in the volume element. 

The coefficients of Terms~(\ref{eq:4.6a})--(\ref{eq:4.6c}) have the form $m^2/2$,
where $m$ is the mass of the tensor. These terms have the same form
as the mass terms for the electroweak vectors, Terms~(\ref{eq:5.7x})--(\ref{eq:5.7z}),
the coefficients of which also have the form $m^2/2$.
Thus we can determine the masses of the tensor particles by taking the ratio of
the numerical part of their coefficients to those of the vectors. For example, the numerical
part of coefficient of the mass term for $g_{\m\n}^{10}$ is
$(2\sqrt 3+\sqrt{30})/12$ [from Term~(\ref{eq:4.6c})]. The numerical part of the coefficient of
the corresponding vector $A_\m^{10}$, which is the $Z^0$,
is $\sqrt{30}/3$ [from Term~(\ref{eq:5.7z})]. Their ratio is
\q
\frac{m^2(g_{\m\n}^{10})}{m^2(Z^0)}=\frac{(2\sqrt3+\sqrt{30})/12}{\sqrt{30}/3},
\qq
where $m^2(g_{\m\n}^{10})$ is the square of the mass of $g_{\m\n}^{10}$
and $m^2(Z^0)$ is the square of the mass of the $Z^0$.
This equation can then easily be solved for the mass of $g_{\m\n}^{10}$
in terms of the mass of the $Z^0$. The result is
\q
m(g_{\m\n}^{10})=0.6388m(Z^0).
\qq

A similar comparison of the tensor particle $g_{\m\n}^{1,\pm1}$
may be made to the corresponding vector with the same quantum numbers
$l$ and $m$, namely $A_\m^{1,\pm1}$, which is the $W^{\pm}$. We have
\q
m(g_{\m\n}^{1,\pm1})=0.4887m(W^{\pm}),
\qq
where $m(g_{\m\n}^{1,\pm1})$ is the mass of $g_{\m\n}^{1,\pm1}$
and $m(W^{\pm})$ is the mass of the $W^{\pm}$.

Much of the error between the theoretical values of masses for the
 $g_{\m\n}^{1m}$ and their potential observed values can be eliminated
simply by substituting the observed masses for the $Z^0$ and $W^\pm$
instead of their values from theory in Sec.~21. This results in the masses of
58.25 GeV for  $g_{\m\n}^{10}$ and 39.29 GeV for  $g_{\m\n}^{1,\pm1}$. 
Apparently, the $g_{\m\n}^{1m}$ are hard to find for the same
reason the so-called Higgs boson was hard to find. The 
background is too high. They have similar decay products.

There arise in the Lagrangian $k\widehat R$ the terms    
\begin{equation}
-k\w g^{\,55}\w g^{\,66}\p_5^{\ 2}\w g_{66}
-\tfrac12\,k\w g^{\,55}\p_5\w g^{\,66}\p_5\w g_{66}.
\label{eq:4.8}
\end{equation}
If one makes the usual choice for the HD coordinates
$\w x^{\,5}=\theta$ and $\w x^{\,6}=\phi$, then
$\w g_{66}=-c^2 T^2\sin^2\!\!\theta$,
$\w g^{\,66}=-1/(c^2 T^2\sin^2\!\!\theta)$ and
$\w g^{\,55}=-1/(c^2 T^2)+A^\mu A_\mu$.
The derivatives of $\w g_{\,66}$ and $\w g^{\,66}$ with respect to $\theta$ 
do not vanish.
Terms~(\ref{eq:4.8}) then contain the term $2k/(c^2 T^2)$,
which is a cosmological constant roughly the size of the Planck mass.
However, with our choice of $\w x^{\,5}$ and $\w x^{\,6}$, the field 
$\w g_{66}=\w g^{\,66}=1$. Therefore, these terms are zero.
Thus, incorporating $\sin\!\theta$ into the coordinate $\w x^{\,6}$
instead of the scalar $\w g_{66}$ eliminates this large
cosmological constant.
\pagebreak[2]

{\samepage
\begin{center}
{\bf 6. The electroweak vectors}
\end{center}

\nopagebreak
In this section we derive Maxwell and mass terms for the photon, W and Z.
The free-$A_\m$-field terms have anywhere from one to six factors of $A$.
Of these, we consider here only terms with one or two of these factors.
These terms are more important and are easier to obtain.  They are
\begin{eqnarray}
+\tfrac{1}{2}k\delta^{\r\s}&\delta^{\m\n}&\left(
\p_\r A_\m\p_\n A_\s
-\p_\r A_\m\p_\s A_\n\right) \label{eq:5.1}\\
+2k&\delta^{\m\n}&\p_\m\p_5A_\n\label{eq:5.2}\\
-2k&\delta^{\m\n}&A_\m\p_5^{\ 2}A_\n\label{eq:5.3}\\
-2k&\delta^{\m\n}&\p_5A_\m\p_5A_\n\label{eq:5.4}\\
-\tfrac{1}{2}k&\delta^{\m\n}&\p_6A_\m\p_6A_\n.
\label{eq:5.4a}
\end{eqnarray}}

A word of warning about this calculation.
Like many calculations in general relativity,
arriving at these terms is a straightforward, but lengthy calculation.
One must devote many hours to it if one is to arrive at the correct result.
One should limit oneself to terms with two or less factors of A
as soon as possible in the calculation. Terms~(\ref{eq:5.1})
are the Kaluza-Klein miracle --- they lead to Maxwell's
equations for the photon when the HD coordinate
dependence of $A_\mu$ is neglected and $\partial_5=\partial_6=0$.
In this case the calculation is much simpler and takes less than an hour.
Terms~(\ref{eq:5.3})--(\ref{eq:5.4a})  are similar
to Terms~(\ref{eq:4.2})--(\ref{eq:4.5}), which are much easier to
obtain. Terms like~(\ref{eq:5.2}), with only one factor of A,
appear only twice in the Lagrangian before they are added together.
Because these terms have one second derivative, one may neglect the
terms within the products $\Gamma\Gamma$, which have two first derivatives,
in the curvature scalar when deriving them.

Terms~(\ref{eq:5.1}) will lead to Maxwell's equations for the photon,
W and Z and predicted intermediate vector bosons.  Term~(\ref{eq:5.2}) will
lead to the Dirac equation for fermions.
Terms~(\ref{eq:5.3})--(\ref{eq:5.4a}) will lead to mass terms
for vectors and interaction terms for spinors with vectors.
Terms with three to six factors of A are self-interaction terms for vectors,
similar to those found in the Standard Model.

The procedures for evaluating these terms by differentiating and integrating
over the spherical harmonics in them are described in the next section.
Because $g^{\,\m\n}=\delta^{\m\n}$, then $\delta^{\m\n}$ acts as the contravariant
4D metric tensor and is associated with the bra $\langle00|$ in these terms. 
Because the vector $A^\m$ is really two separate fields,
$\delta^{\m\n}$ and $A_\n$, it should be written as $\delta^{\m\n}A_\n$
when integrating over $\theta$ and $\phi$; otherwise the result of
integration will be incorrect. This is explained in the next section.
The expansion for $A_\m$ is carried to $l=1$ for now.  

Terms~(\ref{eq:5.1}) are the only terms
with two factors of $A$ and two 4D derivatives.
The result for these terms is 
\q
-\tfrac{k}{4}\,
F^{\m\n}_{00}F_{\m\n}^{00}
-\tfrac{k}{4\sqrt3}\,
F^{\m\n}_{1,-1}F_{\m\n}^{11}
+\tfrac{k}{4\sqrt3}\,
F^{\m\n}_{10}F_{\m\n}^{10}
-\tfrac{k}{4\sqrt3}\,
F^{\m\n}_{11}F_{\m\n}^{1,-1},\label{eq:5.5}
\qq
where $F_{\m\n}^{lm}=\p_\m A_\n^{lm}-\p_\n A_\m^{\m\n}$.
We will identify $A_\m^{00}$ with the photon field shortly.

We introduce the reality condition 
\q
\sum A^\m_{lm}Y_{lm}=(\sum A^\m_{lm}Y_{lm})^* \label{eq:5.5z}
\qq
and equate coefficients of like $Y_{lm}$ on both sides of this equation.
This produces the relation 
\q
\overline A^{\,\m}_{lm}=(-1)^m A^\m_{l,-m},\label{eq:5.5x}
\qq
where no sum over $m$ is implied.
Further, we introduce the convention of representing the contravariant fields
$A^\m_{lm}$ and $F^{\m\n}_{lm}$ as the complex conjugates $\overline A^{\,\m}_{lm}$ 
and $\overline F^{\,\m\n}_{lm}$. Terms~(\ref{eq:5.5}) now become
\q
-\tfrac{k}{4}\,
\overline F^{\,\m\n}_{00}F_{\m\n}^{00}
+\tfrac{k}{4\sqrt3}\,
\overline F^{\,\m\n}_{11}F_{\m\n}^{11}
+\tfrac{k}{4\sqrt3}\,
\overline F^{\,\m\n}_{10}F_{\m\n}^{10}
+\tfrac{k}{4\sqrt3}\,
\overline F^{\,\m\n}_{1,-1}F_{\m\n}^{1,-1},\label{eq:5.5a}
\qq
where $\overline F^{\,\m\n}_{lm}=\p^\m \overline A^{\,\n}_{lm}-\p^\n \overline A^{\,\m}_{\m\n}$.

The signs of the Maxwell terms for $A_\m^{11}$, $A_\m^{1,-1}$, and $A_\m^{10}$
are opposite what they should be. 
This will also be the case for their mass terms.
This means these fields have negative energy. However, this is no more of a problem 
than the existence of the negative energy positron, which follows from the 
Dirac equation. Both problems are dealt with as follows.
Negative energy states were thought to be unstable
because there would be no ``barrier" preventing them from 
decaying into negative energy states with increasingly lower energy.
However, if the ground state
is defined to be the state closest to zero energy instead
of lowest energy, then this cannot happen. Particles with negative energy
tend to gain energy, not lose it. We adopt this 
convention so that there is a symmetry around zero energy.
Note that the term for the photon has positive energy. 

The fields are now redefined to yield the conventional coefficient
of the Maxwell term for the photon
\begin{equation}
A_\m^{\pp lm}=\left(\frac{c^4}{16\pi G}
\right)^{1/2}A_\m^{lm},\label{eq:5.6}
\end{equation}
where we have included the factor of $c$ from $x^4=ict$ in the
differential volume element in the redefinition.
Terms~(\ref{eq:5.5a}) now become
\q
-\tfrac{1}{4}\,
\overline F^{\,\pp\m\n}_{00}F_{\m\n}^{\,\pp00}
+\tfrac{1}{4\sqrt3}\,\overline F^{\,\pp\m\n}_{11}
F_{\m\n}^{\,\pp11}
+\tfrac{1}{4\sqrt3}\,
\overline F^{\,\pp\m\n}_{10}F_{\m\n}^{\,\pp10}
+\tfrac{1}{4\sqrt3}\,
\overline F^{\,\pp\m\n}_{1,-1}
F_{\m\n}^{\pp1,-1},\label{eq:5.5'}
\qq
where $F_{\m\n}^{\pp lm}=
\p_\m A_\n^{\pp lm}
-\p_\n A_\m^{\pp lm}$.
Henceforth, the redefinition of $A_\m^{\pp lm}$ will be understood and
we will drop their prime.
These terms appear in the Lagrangian for the SM.
They are the free-field (Maxwell) terms for the photon,
W and Z.  The photon is identified with $A_\mu^{00}$;
the $W_\mu^-$ is $A_\mu^{1,-1}$,
the $Z_\mu^0$ is $A_\mu^{10}$,
and the $W_\mu^+$ is $A_\mu^{11}$.

We have
\begin{eqnarray}
\w g_{\mu5}&=&A_\mu\label{1}\\
           &=&\sum_{l=0}^{\infty}\sum_{m=-l}^{l}A_\mu^{lm}Y_{lm}(\theta,\phi)\\
           &=&A_\mu^{00}\,Y_{00}+A_\mu^{1,-1}\,Y_{1,-1}+A_\mu^{10}\,Y_{10}
            +A_\mu^{11}\,Y_{11}+\ldots\\
           &=&A_\mu^{00}\,Y_{00}+W_\mu^{-}\,Y_{1,-1}+Z_\mu^{0}\,Y_{10}
            +W_\mu^{+}\,Y_{11}+\ldots
\end{eqnarray}
The photon is denoted by $A_\mu^{00}$, so as not to be confused with
$A_\mu$, which is given in Eq.\,(\ref{1}).  We use only the physical fields;
there is no auxiliary vector ${\bf B_\mu}$ as found in the SM, nor subsequent mixing.
Note that we have given the W and Z the same normalization as the
photon. This is the simplest thing to do and will be absolutely necessary
in order to obtain the masses of the W and Z from the coefficients
of their mass terms. This normalization for the W and Z leaves a factor
of $1/\sqrt3$ in their Maxwell terms. This does not
contradict observation as it would for the photon because one does
not detect the W and Z directly; one can only observe their decay
products. 

The result for Term~(\ref{eq:5.2}) is
$\tfrac{\sqrt 3}{2}\,\pi(cT)^{\ -1}k\delta^{\m\n}\p_\m A_\n^{10}$.
This term is a divergence, which can be transformed into a
surface integral at infinity, where the fields vanish. Therefore,
Term~(\ref{eq:5.2}) is zero for the $A_\m^{lm}$.

Terms~(\ref{eq:5.3})--(\ref{eq:5.4a}) are the only ones 
with two factors of A and two HD derivatives.  
The result for these terms is
\begin{eqnarray}
-\frac{9+4\sqrt3\,}{12c^2T^2}\,
&\delta^{\m\n}&A_\m^{1,-1}
A_\n^{11}\label{eq:5.7a}\\
-\frac{9+4\sqrt3\,}{12c^2T^2}\,
&\delta^{\m\n}&A_\m^{11}
A_\n^{1,-1}\label{eq:5.7b}\\
+\frac{\sqrt{30}\,}{3c^2T^2}\,
&\delta^{\m\n}&A_\m^{10}
A_\n^{10}.\label{eq:5.7c}
\end{eqnarray}
We employ the above-mentioned 
convention of representing the first factor of $A$ in each term as a 
contravariant vector whose complex conjugate is taken. 
The result for these terms is now
\begin{eqnarray}
+\frac{9+4\sqrt3\,}{12c^2T^2}\,
&\overline A^{\,\m}_{11}
A_\m^{11}\label{eq:5.7x}\\
+\frac{9+4\sqrt3\,}{12c^2T^2}\,
&\overline A^{\,\m}_{1,-1}
A_\m^{1,-1}\label{eq:5.7y}\\
+\frac{\sqrt{30}\,}{3c^2T^2}\,
&\overline A^{\,\m}_{10}
A_\m^{10}.\label{eq:5.7z}
\end{eqnarray}
These are mass terms for the W and Z.  Like
$g_{\m\n}^{00}$, the field $A_\m^{00}$ does not have a
mass term because its associated spherical harmonic $Y_{00}=1/\sqrt{4\pi}$ 
does not depend on the HD coordinates. Therefore
$A_\m^{00}$ is massless, which is consistent with our
identification of it as the photon.

It is a miracle that the 6D curvature scalar $\w R$ leads to the 4D 
curvature scalar $R$ plus the Maxwell term $-\frac{1}{4}F^{\m\n}F_{\m\n}$
when fields do not depend on the HD coordinates $\w x^5$ and $\w x^6$.
The calculation is simple and seems to unify gravitation with electromagnetism.
It is miraculous that so little
effort leads to such a good result. However, there is a fly in the ointment.
There is no reason why fields should not depend on the HD coordinates.
If general relativity is truly six-dimensional, fields should depend on all
six coordinates, not just the first four. After all, this is how four-dimensional general
relativity works. The fields depend on all four coordinates in this case.
Therefore, we will make our theory more correct by allowing fields to depend on 
all six coordinates. This allows the miracle to continue.

Following from the HD coordinate dependence of $A_\m$ is
the existence of the W and Z. Following from terms with two 4D and two HD derivatives
are the terms that describe them. Orthogonality of the
spherical harmonics produces one Maxwell and one mass term for each
of the W's and Z when the Lagrangian density is integrated over
$\theta$ and $\phi$. Terms with three to six factors of $A_\mu$
lead to self-interaction terms for the W and Z similar to the 
Yang-Mills self-interaction terms found in the SM. These results are 
examples of the details following from properly fixed postulates.

The theory of 3D time explains why there are three weak vectors
and one electromagnetic vector. It explains why the weak vectors have
charges $-1$, 0, $+1$. It explains why the photon has no charge.
(Charge is the momentum canonically conjugate to the coordinate dependence
$e^{i\phi}$.) It explains why the W and Z have mass while the photon does not.
This derivation of the existence of the photon,
W and Z, and their properties explains the origin of what appears to be the local
symmetry group $SU(2)\times U(1)$, which must be assumed in the Standard Model.
We have produced all necessary terms for these fields using spherical harmonics
and the curvature scalar as Lagrangian density --- both natural parts
of general relativity with three dimensions of time.
\pagebreak[2]

{\samepage
\begin{center}
{\bf 7. How to differentiate and integrate
over the $Y_{lm}(\theta, \phi)$}
\end{center}

\nopagebreak
This section may be omitted in a first reading.  It describes how to
go from `raw' terms in the Lagrangian such as 
Terms~(\ref{eq:5.1})--(\ref{eq:5.4a}) to their more final forms such as
Terms~(\ref{eq:5.7a})--(\ref{eq:5.7c}). 
Without an organized procedure for this, one soon runs into
expressions that cannot be evaluated because of infinities.
In addition, there will be more than one way to do certain calculations,
each with a different result. The right way must be specified.}

If one attempts to multiply three or more expansions in terms of
spherical harmonics found in terms with three or more factors of $A$
in the Lagrangian, one soon runs into a mess consisting of too many
factors of $\sqrt\pi$ as well as a plethora of other square roots.
These do not all disappear after the term is integrated over
$\theta$ and $\phi$ as they do in terms with only two expansions.
There must be a better way.  Indeed there is:  One must convert
the spherical harmonics to kets.
Then products of two kets may be written as single
kets using Clebsch-Gordan coefficients.  This process is continued
until there is just one ket for the contravariant group of factors and
one ket for the covariant group. Then the ket for the contravariant group
is written as a bra and the final bra and ket are multiplied
together representing the integral over $\theta$ and $\phi$.
Or, if one chooses, the final bra and ket may be reconverted
to spherical harmonics.  Because we are integrating here over just two
spherical harmonics, the excess factors of $\sqrt\pi$ and
other square roots appearing in each spherical harmonic will not appear
in the answer.  Thus, we will have converted integrals over three 
or more spherical harmonics into those over just two.  

Before the procedures for taking the higher-dimensional derivatives
and integrating over the spherical harmonics are discussed, I will
show how to deal with the bras and kets in each term. 
As described in Sec.\,4, $\gggm$ is the 4D
contravariant metric tensor and
raises 4D indices. Therefore, we have equations like
\begin{equation}
A_{lm}^{\m}=g^{\,\m\n}_{00}A_{\n}^{lm}.
\label{eq:A1}
\end{equation}
This shows that $g_{00}^{\,\m\n}$ is the contravariant
part of $A_{lm}^{\m}$. Since every `contravariant'
tensor may be written in this
fashion, then $\gggm$ is the only true contravariant tensor.
It matters whether $A_{lm}^{\m}$ or $\gggm$ is the
contravariant tensor
because the results of integrating over $\theta$ and $\phi$
depend upon which is chosen. For example, if $A_{lm}^{\m}$ 
is the contravariant tensor, we would have
\begin{equation}
\overline A_{11}^{\,\m}\langle11|A_{\m}^{11}|11\rangle
=\overline A_{11}^{\,\m}A_{\m}^{11},\label{eq:A2}
\end{equation}
while if $\gggm$ is the contravariant tensor, then the same expression
could be written as 
\begin{equation}
-g_{00}^{\,\m\n}\langle00|A_{\m}^{1,-1}
|1,-1\rangle A_{\n}^{11}|11\rangle
=\tfrac{1}{\sqrt3}\,\overline A_{11}^{\,\m}A_{\m}^{11},\label{eq:A3}
\end{equation}
where the kets $|1,-1\rangle$ and $|11\rangle$ are combined first.
There is a factor of $\sqrt3$ difference between these expressions.
Equation~(\ref{eq:A3}) is correct because $\gggm$ is the only
true contravariant tensor.

Because factors are expanded in terms
of spherical harmonics, which are then converted into kets,
which do not commute,
factors in a term must be ordered properly.  This order
is determined by the above definition of the curvature scalar
in terms of the metric tensor. We will be dealing with terms
with no more than two nontrivial covariant factors. Therefore the order
of the two covariant factors does not matter because integration with
$\langle 00|$, which is the only bra we will be using, and two kets
acts like a dot product between the two kets.

As the kets are nonassociative, the order of the multiplication
of factors in a term must be specified.  This order is largely
implied by the structure of the equations that define the
curvature scalar.  First, the products of the factors of $A$ in each of
Eqs.\,(\ref{eq:2.8a}) and~(\ref{eq:2.8b}) are taken.  
Then the factors within
$\w\G^{\,\dd}_{\ \a\bb}$, $\w R_{\a\bb}$ and
$\w R$ are combined in that order.
In practice, these rules are not necessary because we will be
dealing here with terms with no more than two nontrivial covariant factors.
This means there is only one nontrivial multiplication,
making the order of multiplications irrelevant.

Note that the order of factors and the order of
multiplications in a term pertain only to the covariant factor
[for example, $A_{\n}^{lm}$ in Eq.\,(\ref{eq:A1})] within the
seemingly contravariant factor [$A_{lm}^{\m}$ in Eq.\,(\ref{eq:A1})].
The contravariant factors $\gggm$ are not tied down to a particular
position in the term nor in the order of multiplications like the
covariant factors, because they are associated with the ket
$|00\rangle$ only.  Therefore, we may group
all of the 4D contravariant metric tensors in a
term together and place them to the left
of the covariant group of factors. The adjoint is then taken of the 
group of $g_{00}^{\,\m\n}$.  The reason the $\gggm$ are grouped 
together is that we will want to combine the kets of the  
contravariant factors and, separately, the kets of the covariant
factors. We then write the adjoint of the resultant ket for the
contravariant factors as a bra. This is so that there will be one bra and
one ket, representing the contravariant and covariant factors,
respectively.  The final bra and ket are then combined, representing
one integral over $\theta$ and $\phi$.  If the 
contravariant factors were not grouped together, we would have
more than one integral over $\theta$ and $\phi$ per term.
The contravariant group of factors is placed to the left
of the covariant group because bras must be placed to the
left of kets if their products are to represent integrals.
Note that the only bra we will ever have to consider for the 
contravariant factors is $\langle 00|$.

Term~(\ref{eq:5.3}) has the factor $\partial_5^{\ 2}$, which equals
$-1/(c^2 T^2)\partial^{\ 2}_\theta$, according to Sec.\,3.
Here $\partial_\theta=\partial/\partial\theta$.
To see how this term is integrated over $\theta$ and $\phi$,
we examine the case where $l=0$ or 1 in the expansion for
$A_{\m}$.  The derivatives with respect to $\theta$
are taken: $\partial^{\ 2}_\theta Y_{00}=0$ and $\partial^{\ 2}_\theta
Y_{1m}=-Y_{1m}$.
If $l>1$, this simplification for $\partial^{\ 2}_\theta Y_{1m}$ is
not possible and the expression for $\partial_\theta$ described in the
procedure for Term~(\ref{eq:5.4}) must be used.

The spherical harmonics of the contravariant and covariant factors
are now written as kets $Y_{lm}\to|lm\rangle$ in preparation for
combining them. Next, the two covariant expansions are 
multiplied and the products of kets are written in terms of
single kets using Clebsch-Gordan coefficients.  For example,
\begin{equation}
|11\rangle|1,-1\rangle=\tfrac{1}{\sqrt6}\,|20\rangle
+\tfrac{1}{\sqrt2}\,|10\rangle+\tfrac{1}{\sqrt3}\,|00\rangle.\label{eq:A4}
\end{equation}
Because the metric tensor must be constant for the free-$A_\mu$ fields,
the special relativity metric
applies to the free-$A_\mu$-field terms, including Term~(\ref{eq:5.3}).
Therefore, we set $g^{\,\mu\nu}_{00}=\delta^{\,\mu\nu}$,
where $\delta^{\,\mu\nu}$ now has the ket $|00\rangle$.
The adjoint of the
contravariant factor is taken and its ket is written as a bra.
The contravariant and covariant expressions are now multiplied
together.  The integrals over $\theta$ and $\phi$ may be 
expressed in terms of the orthogonality of the kets:
$\langle00|lm\rangle=\delta_{0l}\delta_{0m}$. 

If one is doing this calculation for $A_\mu^{lm}$ with $l>1$, 
factors of $e^{\pm2i\phi}$ must be combined with the kets
of the differentiated factor before the kets of the two covariant
factors are combined. For example, $e^{2i\phi}|2,-1\rangle=-|21\rangle$.
Otherwise, a different result is obtained.
This procedure must be correct because it leads to the equation
$\partial_\theta^{\ 2}Y_{1m}
=-Y_{1m}$ for the case $l=1$. (Note, however, that $e^{2i\phi}|2,-2
\rangle$ cannot be combined preliminarily because it does not equal
another ket.)

Term~(\ref{eq:5.4}) contains two factors of $\partial_5$. It
is dealt with in the same way as Term~(\ref{eq:5.3}),
except that $\partial_\theta$ is rewritten
\begin{equation}
\partial_\theta=\frac{1}{2\hbar}\left(e^{-i\phi}L_+-e^{i\phi}L_-\right),
\label{eq:A5}
\end{equation}
where $L_+$ and $L_-$ are the raising and lowering operators,
respectively, for the spherical harmonics.  We have
\begin{equation}
L_\pm Y_{lm}=\left[l(l+1)-m(m\pm1)\right]^{1/2}\hbar Y_{l,m\pm1}.
\label{eq:A6}
\end{equation}
This definition for $\partial_\theta$ follows from combining
$L_+=L_x+iL_y$ and $L_-=L_x-iL_y$, where
\begin{eqnarray}
         L_x&=&i\hbar(\sin\!\phi\,\partial_\theta
             +\cot\!\theta\cos\!\phi\,\partial_\phi),\label{eq:A7a}\\
         L_y&=&i\hbar(-\cos\!\phi\,\partial_\theta
             +\cot\!\theta\sin\!\phi\,\partial_\phi),\label{eq:A7b}
\end{eqnarray}
which are the $x$ and $y$ components of the angular momentum vector
${\bf L}$ written in spherical coordinates.
Rewriting the operator $\partial_\theta$ is necessary to write
$\partial_\theta Y_{lm}$ properly in terms of spherical harmonics,
which can then be written as kets.

When one uses this expression for $\partial_\theta$, one winds up
with factors of $e^{i\phi}$ in the term. In order to evaluate
these expressions, the terms have their final bra and ket
rewritten as spherical harmonics according to the transformations
$\langle00|\to Y_{00}$ and $|lm\rangle\to Y_{lm}$ and the terms
are explicitly integrated over $\theta$ and $\phi$ in the usual
way. Note that the factors of $e^{i\phi}$ must cancel if the integral
over $\phi$ and the term is to be nonzero.
For terms without functions of $\phi$, use may be made of
the orthonormality of the kets.

Term~(\ref{eq:5.4a}) contains two factors of $\partial_6$, where
$\partial_6=\partial/(icT\sin\!\theta\partial\phi)$ from Sec.\,3.
This term is dealt with in the same way as Term~(\ref{eq:5.3}).
First, the derivatives with respect to $\phi$ of the $Y_{lm}$ are taken
$\partial_\phi Y_{lm}=imY_{lm}$. In order to evaluate these terms with two
factors of $1/\sin\!\theta$ from the two factors of $\partial_6$,
they have the spherical harmonics of both covariant
factors written in terms of spherical 
harmonics with their $l$ and $|m|$ values decreased by one. (The
quantum number $m$ is increased by one if it is negative.)
Some examples of this are
\begin{eqnarray}
Y_{1,\pm1}&=&\mp\sqrt{\tfrac32}\,Y_{00}\sin\!\theta e^{\pm i\phi},
\label{eq:A8a}\\
Y_{2,\pm2}&=&\mp\sqrt{\tfrac54}\,Y_{1,\pm1}\sin\!\theta e^{\pm i\phi},
\label{eq:A8b}\\
Y_{3,\pm1}&=&\mp\sqrt{\tfrac{7}{12}}\left(\sqrt5\,Y_{20}+Y_{00}\right)
\sin\!\theta e^{\pm i\phi}.\label{eq:A8c}
\end{eqnarray}
The factors of $\sin\!\theta$ in the $\partial_6$'s cancel with those
extracted from the covariant factors in
Eqs.\,(\ref{eq:A8a})--(\ref{eq:A8c}).
This eliminates the factor of $1/\sin^2\theta$,
which would otherwise result in infinity when integrating over $\theta$.
The spherical harmonics are now written as kets and combined.
\pagebreak[2]

{\samepage
\begin{center}
{\bf 8. Leptons}
\end{center}

\nopagebreak
Now that we have the four electroweak vectors, fermions are not far behind.
They follow from Chapter 41 of Misner, Thorne and Wheeler [45], the
classic textbook. This book details standard spinor theory. It states that any
vector is equivalent to a second rank spinor. One does not have to invent a
fantastic theory like supersymmetry to obtain fermions. If one starts with the
right postulates, they come from a textbook.

For any vector $W_\m$, we have
\begin{equation}
W_\m=-\h\pa w_{B \dot V},\label{eq:6.1}
\end{equation}
where $w_{B\dot V}$ is a second rank spinor.  A sum is implied over the
indices $B$ and $\dot V$.  These indices take the values
1, 2 and $\dot 1$, $\dot 2$, respectively.  The 
$\oo\s^{B\dot V}$ are the Pauli spin matrices and
$\s_0^{B\dot V}$ is the unit matrix. These matrices are the basis vectors for
an expansion of a vector. The spinor field components are its coefficients.}

When the vector is null, like for example, the photon, the
second rank spinor is equated to the product of
two 2-component spinors. This produces the product form $\overline \psi \psi$ of spinors necessary
to reproduce the Lagrangian for the Dirac equation.
For example, for the photon we have
\q
\am|00\rangle=-\h\pa\xi_B\eta_{\dot V}
\tfrac{1}{\sqrt2}\k\kk
+\h\pa\tau_B\omega_\wdv\tfrac{1}{\sqrt2}\kk\k,\label{eq:6.2}
\qq
where
\begin{equation}
|00\rangle=\tfrac{1}{\sqrt2\,}\k\kk
            -\tfrac{1}{\sqrt2\,}\kk\k.\label{eq:6.2a}
\end{equation}
The spinor $\xi_B$ or $\omega_\wdv$ has the ket $\k$, while $\eta_\wdv$ or
$\tau_B$ is in the state $\kk$. By convention, the first ket in each pair in
Eq.\,(\ref{eq:6.2}) is identified with the spinor with an undotted index,
while the second is matched with the dotted-index spinor. We have
\begin{equation}
\am=-\h\pa\xi_B\eta_\wdv\label{eq:6.2b}
\end{equation}
for the first pair of kets and
\begin{equation}
\am=-\h\pa\tau_B\omega_\wdv\label{eq:6.2c}
\end{equation}
for the second.

The spinors in this expansion may recombine to form $A_\m^{10}$ (the $Z^0$).
We have
\q
A_\m^{10}|10\rangle=-\h\pa\xi_B\eta_{\dot V}
\tfrac{1}{\sqrt2}\k\kk
-\h\pa\tau_B\omega_\wdv\tfrac{1}{\sqrt2}\kk\k,\label{eq:6.3}
\qq
where
\begin{equation}
|10\rangle=\tfrac{1}{\sqrt2\,}\k\kk
            +\tfrac{1}{\sqrt2\,}\kk\k.\label{eq:6.3a}
\end{equation}

The derivative $\partial_5$ in Term~(\ref{eq:5.2}), which will lead to the
Dirac equation, eliminates $A_\mu^{00}|00\rangle$ from the expansion of
$A_\mu$. The factors of $e^{i\phi}$ in the spherical harmonics for 
$A_\mu^{1,\pm 1}$ eliminate them upon integration over 
$\theta$ and $\phi$. This leaves $A_\mu^{10}|10\rangle$.
After substituting this only nonzero term into Term~(\ref{eq:5.2}), 
the spinor kets are converted to $|10\rangle$ and $|00\rangle$ (for example, 
$\kk\k=\tfrac{1}{\sqrt2}|10\rangle-\tfrac{1}{\sqrt2}|00\rangle$). This results in
\begin{equation}
\frac{\sqrt{3}\,\pi ik}{8cT}\delta^{\m\n}\pa(\xi_B\partial_\n\eta_\wdv
+\partial_\n\xi_B\eta_\wdv),\label{eq:6.5}
\end{equation}
plus a similar set of terms involving the spinors $\tau_B$ and
$\omega_\wdv$.  The spinors $\eta_{\dot V}$ and $\xi_B$ will be
identified with the electron field and its complex conjugate, respectively,
while $\omega_\wdv$ and $\tau_B$ will be the neutrino and its
conjugate, respectively. These identifications are suggested by the
associations these fields have with their kets above, according to
their quantum numbers $l$ and $m$ given in Table~1.

We will ignore the neutrino terms for now and concentrate
on deriving the Dirac equation for the electron.
The fields are redefined to give them the right units
\begin{equation}
\xi^{\,\prime}_B=\left(\frac{\sqrt3\,c^3}{128\hbar GcT}
\right)^{1/2}\xi_B,\label{eq:6.6}
\end{equation}
with an identical redefinition for $\eta_\wdv$. 
The result for Term~(\ref{eq:5.2}) is now
\begin{eqnarray}
+&i\hbar c&\delta^{\m\n}\pa\xi^{\,\prime}_B\partial_\n
\eta^{\,\prime}_\wdv\label{eq:6.7a}\\
+&i\hbar c&\delta^{\m\n}\pa\partial_\n\xi^{\,\prime}_B\eta^{\,\prime}_\wdv.\label{eq:6.7b}
\end{eqnarray}

Spinor mass terms are not contained directly within the 6D curvature scalar.
One must introduce spinors into the Lagrangian
with the above-mentioned substitution for vectors, which does not
allow terms of the form $-m\xi^{\,\prime}_B\xi^{\,\prime B}$.
However, new terms may result from a particle's
motion on the HD sphere. This motion produces energy and must 
be represented in the Lagrangian. 
Motion on the HD sphere is isospin. Electric charge will be related to 
the $z$-component of isospin in Part~II. Further,
spinors acquire mass from their charge according to
the classical radius equation considered in Part~II. 
Therefore, we introduce the mass terms
\begin{equation}
-m\xi^{\,\prime}_B\xi^{\,\prime B}-m\eta^{\,\prime\wdv}
\eta^{\,\prime}_\wdv.\label{eq:6.8}
\end{equation}
Although these mass terms are not contained directly within $\w R$,
they do not produce anomalies because they are derived from $\w R$.
It is $\w R$ that puts a moving electron on the HD sphere.
This is indicated by the electron's HD angular momentum, which is
given by the operators $L_z$ and $L^2$ operating on its ket.

Each of Terms~(\ref{eq:6.7a}) or~(\ref{eq:6.7b}) leads to the
Dirac equation. We will
consider only Term~(\ref{eq:6.7a}) in what follows.
Lagrange's equations for $\xi'_B$ and $\eta'_\wdv$ are
\begin{eqnarray}
&i\hbar c&\delta^{\m\n}\pa\partial_\n\xi'_B=m\eta^{\,\prime\wdv},\label{eq:6.9a}\\
&i\hbar c&\delta^{\m\n}\pa\partial_\n\eta'_\wdv=m\xi^{\,\prime B}.\label{eq:6.9b}
\end{eqnarray}
Using the equations [45] $\zeta^B\xi'_B=-\zeta_B\xi^{\,\prime B}$
and $\eta^{\,\prime\wdv}=\varepsilon^{\dot V\dot W}\eta'_{\dot W}$,
where the only nonzero components of $\varepsilon^{\dot V\dot W}$
are $\varepsilon^{\dot1\dot2}=-\varepsilon^{\dot2\dot1}=1$,
Eq.\,(\ref{eq:6.9a}) becomes
\begin{equation}
-i\hbar c\s^\m_{B\dot V}\partial_\m\xi^{\,\prime B}
=m\eta'_\wdv.\label{eq:6.9a'}
\end{equation}
Here the $\s^\m_{B\dot V}$ are the associated basic spin matrices.
They are computed from the $\pa$ via the relations
\begin{equation}
\s^\m_{B\dot V}\s_\n^{B\dot V}=-2\delta^\m_{\ \n}.\label{eq:6.10}
\end{equation}
The $\s^\m_{B\dot V}$ are equal to [45] minus the $\pa$, except for
$\s^2_{B\dot V}$, which equals $\s_2^{B\dot V}$.

Equations~(\ref{eq:6.9a'}) and~(\ref{eq:6.9b}) may be put in matrix form
\begin{eqnarray}
\left(i\partial_0+i\oo\sigma\cdot\oo\nabla\right)\xi&=&m\eta,
     \label{eq:6.11z}\\
\left(i\partial_0-i\oo\sigma\cdot\oo\nabla\right)\eta&=&m\xi,\label{eq:6.11}
\end{eqnarray}
where the $\oo\s$ are the Pauli spin matrices and
\begin{equation}
\xi=\left(\begin{array}{c}
\xi^{\,\prime1}\\ 
\xi^{\,\prime2}\end{array}\right),
\qquad\eta=\left(
\begin{array}{c}\eta^{\,\prime}_{\dot1_{}}\\ 
\eta^{\,\prime}_{\dot2}
\end{array}\right).\label{eq:6.11a}
\end{equation}
We have set $\hbar=c=1$, which applies for the remainder of this
section. Adding and subtracting Eqs.\,(\ref{eq:6.11z}) and~(\ref{eq:6.11}),
we obtain [58]
\begin{eqnarray}
         i\partial_0\varphi+i\oo\s\cdot\oo\nabla\chi
         &=&m\varphi,\label{eq:6.111'}\\
        -i\partial_0\chi-i\oo\s\cdot\oo\nabla\varphi
         &=&m\chi,
\label{eq:6.11'}
\end{eqnarray}
where
$\varphi=\frac{1}{\sqrt2\,}(\xi+\eta)$ and
$\chi=\frac{1}{\sqrt2\,}(\xi-\eta)$. 

Equations~(\ref{eq:6.111'}) and~(\ref{eq:6.11'}) may be recast as
\begin{equation}
i\gamma^\m\partial_\m\psi=m\psi,\label{eq:6.11''}
\end{equation}
where
\begin{equation}
\psi=\left(\begin{array}{c}\varphi\\ \chi\end{array}
\right),\label{eq:6.11''a}
\end{equation}
\begin{equation}
\gamma^0=\left(\begin{array}{cc}1&0\\0&-1\end{array}\right),\qquad
\oo\gamma=\left(\begin{array}{cc}0&\oo\s\\-\oo\s&0\end{array}\right),
\label{eq:6.11''b}
\end{equation}
where each element of the matrices in Eqs.\,(\ref{eq:6.11''b}) is itself a
$2\times2$ matrix.
Eq.\,(\ref{eq:6.11''}) with~(\ref{eq:6.11''a})
and~(\ref{eq:6.11''b}) constitute the Dirac
equation for the electron in the Dirac representation for $\psi$.

A similar derivation of the Dirac equation for the neutrino
involving the spinors $\tau_B$ and $\omega_\wdv$ may be made.
In addition, the photon may be expanded in terms of spinors with $l>\h$.
The muon and its neutrino are associated with the kets
$\llle\frac32,-\h\rr$ and $\llle\frac32\,\h\rr$, respectively.
The tau and its neutrino are in the respective states
$\llle\frac52,-\h\rr$ and $\llle\frac52\,\h\rr$.
Continuing in this manner, 3D time predicts an infinite
number of fermions, each corresponding to a state $\llle lm\rr$,
where $l$ and $m$ are half-odd integral. The reason why the number of
generations of leptons appears to be limited to three is explained in Part~II.
Thus the expansion of $\w g_{5\mu}=A_\mu$ in terms of spherical harmonics
produces multiple generations of leptons. 

I claim that the fact that the positron appears to have negative energy
is not merely appearance but is deep-rooted and cannot be
transformed away.  The negative energy of the positron is dealt with
in the same way as that for the $Z^0$: The ground state is defined
to be that closest to zero energy instead of lowest energy.

In order to reproduce the SM fully, one must produce the chirality
of the weak interactions.  Chirality is allowed in
3D time because it has an even number of dimensions.
A massless particle must have its right-handed field equal to zero.
This is because one cannot slow down a left-handed particle
and reverse its direction if it travels at the speed of light.
The nearly zero mass of the neutrino is derived in Part~II.
Therefore, $\nu_R=0$ and $\nu=\nu_L$ because $\psi=\psi_L+\psi_R$.
This prevents an interaction term involving right-handed neutrinos,
the W and electrons, producing the required chirality.

The Dirac equation originates from the 6D curvature scalar of 3D time.
This occurs because the divergence
$\partial_\mu A^\mu_{10}$ splits into terms of the form of the Dirac
term $\overline\psi\,\gamma^\mu\partial_\mu\psi$ when the spinor form
$\overline\psi\gamma^\mu\psi$ of the vector $A^\mu_{10}$ is used.

One may wonder why the Maxwell term does not apply to spinors.
If a vector is equal to its spinor equivalent, one could substitute its
spinor form into the Maxwell Terms~(\ref{eq:5.1}). However, examination of
Terms~(\ref{eq:5.1}) and~(\ref{eq:5.2}) reveals that the Dirac term,
which contains $\partial_5$, is larger by a factor $1/(cT)$.
Therefore, the spinor form of the Maxwell term is negligible
compared to the Dirac term.

We have expanded each $A_\m^{lm}$ in terms of the Pauli matrices just like
we expanded $A_\m$ in terms of spherical harmonics. This was made possible 
by the existence of the vectors $A_\m^{lm}$, each of which may be equated to the product
of two 2-component spinors. This is another example of the details following from
properly fixed postulates.

Our method of introducing fermions into a purely bosonic Kaluza-Klein theory 
has several advantages over supersymmetry. Substituting two spinors for each
vector in the theory relates the known spinors to the known vectors. This is highly 
economical and efficient. It does not introduce entire classes of supersymmetric particles for which 
there is little evidence. And it is consistent with our program of keeping the curvature scalar
as Lagrangian and therefore pure gravitation as the theory. Finally, one cannot derive
the Dirac equation from supersymmetry. It must be assumed, like supersymmetry itself.
\pagebreak[2]

{\samepage
\begin{center}
{\bf 9. Interaction terms}
\end{center}

\nopagebreak
Interaction terms for intermediate vector bosons
and leptons may be derived from the
vectors' mass Terms~(\ref{eq:5.7a})--(\ref{eq:5.7c}).
This is done by expanding
one of the vectors in these terms in terms of spinors. 
For example, Term~(\ref{eq:5.7c}) starts off as
\begin{equation}
+\ \ \frac{\sqrt{30}}{3c^2 T^2}\,k\delta^{\mu\nu}A_\mu^{10}A_\nu^{10}.
\label{eq:7.1}
\end{equation}
This term ends up as
\begin{equation}
-\frac83\,\left(\frac{10}{\pi}\right)^{1/2}\sqrt{\hbar c}\,Z_\mu^0\left(
  \overline e\,\gamma^\mu e+{\overline\nu}_e\gamma^\mu\nu_e\right),
  \label{eq:7.1'}
\end{equation}
Thus, the bare weak charge
of the electron or neutrino interacting with the $Z^0$ from the 
6D curvature scalar is
\begin{equation}
g_{\widehat R}=-\frac83\left(\frac{10}{\pi}\right)^{1/2}\sqrt{\hbar c}\,
              =-4.75766\sqrt{\hbar c}\,.\label{eq:7.3}
\end{equation}}
To see where this coefficient comes from, we write each of the factors
that go into its formation. The coefficient is equal to
\begin{equation}
\left(\frac{\sqrt{30}}{3c^2 T^2}\,\right)_1
\left(\frac{c^3}{16\pi G}\right)_2
\left(c\right)_3
\left(\frac{16\pi G}{c^4}\right)_4^{\ 1/2}
\left(\frac{128\hbar GcT}{\sqrt3\,c^3}\right)_5
\left(-\frac12\right)_6
\left(\frac12\right)_7,\label{eq:7.3'}
\end{equation}
where the factors have been numbered for future reference.

The 1st factor is the coefficient of the Z mass term from Sec.\,6.
We note that $cT=L=(\hbar G/c^3)^{1/2}$ is the
Planck length. Factor 2 is the constant $k$ from Sec.\,4, by which
the entire Lagrangian is multiplied. This gives the Lagrangian the
units of energy. The 3rd factor comes from the relativistic
time coordinate $dx^4=icdt$ in the differential volume element of the
action integral for the term.
Factor 4 results from the redefinition of one vector field in
the term described in Sec.\,6. Factor 5 is from the redefinition
of two spinors in the Dirac term in Sec.\,8.
Factor 6 is from Eq.\,(\ref{eq:6.1}), which equates vectors
to spinors. The 7th factor arrises when $|10\rangle$ is
written in terms of $l=1/2$ kets, each pair of which are
multiplied by the Clebsch-Gordan coefficient $1/\sqrt2$. This
Clebsch-Gordan coefficient is squared when the spinors' $l=1/2$ kets
are reconverted to the kets $|10\rangle$ and $|00\rangle$, which is done
to combine them with the bra $\langle 00|$. Another way
of looking at Factor 7 is that the  $Z^0$ is composed of
one-half an electron-positron pair
and one-half a neutrino-antineutrino pair.

We now consider interaction terms involving the muon $\mu^-$.
The Clebsch-Gordan coefficient that relates $|10\rangle$ to the
muon-antimuon's $l=3/2$ pair of spinor kets is $1/\sqrt{20}$.
This means $A_\mu^{10}$ (the Z) may also be considered to be 1/20th a
muon-antimuon pair. This produces a factor of 1/20 in Factor 7 in
Eq.\,(\ref{eq:7.3'}) instead of the factor of 1/2 for the electron.
Thus, the coefficient of the weak interaction term from $\widehat R$
for the muon would appear to be 1/10th that of the electron.
Because the above-mentioned Clebsch-Gordan coefficient for the
muon is 1/10th that of the electron in the Dirac term as well,
the resulting Dirac equation
for the muon would have a factor of 1/10 as coefficient for
the Dirac term as well as the weak-interaction term. Since, at
this point, these are the only two terms in the Dirac equation
for the muon, one may multiply the entire equation by 10 and
recover the Dirac term with coefficient one as well as a weak
charge that is the same as that for the electron. 
The weak charge of the tau from the 6D curvature scalar
remains the same as that for the electron by the same reasoning
used for the muon.

Spinors may combine to form the $W^+$ and $W^-$
\begin{eqnarray}
A_\mu^{11}|11\rangle&=&-\tfrac12\sigma_\mu^{B\dot V}
         \xi_B\omega_{\dot V}\k\k,\label{eq:7.4a}\\
         A_\mu^{1,-1}|1,-1\rangle&=&-\tfrac12\sigma_\mu^{B\dot V}
         \tau_B\eta_{\dot V}\kk\kk,\label{eq:7.4b}
\end{eqnarray}
which, when substituted into Terms~(\ref{eq:5.7a})
and~(\ref{eq:5.7b}), yield
\begin{eqnarray}
+\ \frac{16+12\sqrt3}{3\sqrt\pi}\,\sqrt{\hbar c}\,& W_\mu^+&
         {\overline\nu}_e\gamma^\mu e\label{eq:7.5a}\\
        +\ \frac{16+12\sqrt3}{3\sqrt\pi}\,\sqrt{\hbar c}\,& W_\mu^-&
         \overline e\,\gamma^\mu\nu_e.\label{eq:7.5b}
\end{eqnarray}

There is no photon-electron interaction term in $\w R$. This term results 
from motion on the HD sphere because it depends on the charge of the electron.
As noted above, additional terms in the Lagrangian may result from this motion.
Therefore, we introduce the interaction term
\q
q A_\mu^{00}\,\overline e\,\gamma^\mu e,\label{eq:7.6}
\qq
where $q$ is the charge on an electron. This is similar to the introduction
of the mass terms for spinors in Sec.\,8.

Interaction terms are an important part of the SM, which
produces them through a covariant derivative used in the Dirac term.
This is done to maintain the symmetry $SU(2)\times U(1)$. In 3D time, however,
we have a different but equally effective method for generating
interaction terms. Here the spinor form $\overline\psi\,\gamma^\mu\psi$
of the vector $A^\mu$ is substituted
into a mass term of the form $A^\mu A_\mu$, creating a term of the
form $\overline\psi\,\gamma^\mu\psi A_\mu$.
One may substitute two spinors for both vectors in terms like (\ref{eq:7.1}).
Four-spinor interaction terms result.  These are negligible compared to
vector-spinor interaction terms because they contain an extra factor of
$cT$ from the additional spinor redefinition.  [See Factor 5 of
Eq.\,(\ref{eq:7.3'}).]  This explains why there are no four-point
Fermi interactions and why the weak interactions are carried by
intermediate vector bosons.
\pagebreak[2]

{\samepage
\begin{center}
{\bf 10. The strong interactions}
\end{center}

\nopagebreak
General relativity can be considered to be the result of local Lorentz transformations
in four dimensions.  But what about Lorentz transformations in six dimensions?
When one makes the transition from four to six dimensions, the Lorentz group requires
nine additional generators to describe a Lorentz transformation.
Each of these additional generators corresponds to a rotation or boost
in a coordinate surface with at least one coordinate $\w x^{\,5}$ or $\w x^{\,6}$.
What force is produced by these nine additional types of
Lorentz transformation?  It is the strong interaction.}

The Lorentz group becomes the rotation group for our coordinates
because coordinates of time multiplied by $i$ are spacelike.
This is reflected in the resulting signature diag.\,(~+~+~+~+~+~+~) of the metric tensor.
Imaginary time allows rotations in surfaces formed by an HD coordinate and a coordinate of 4D
spacetime. It is these rotations that act as the source of the strong interactions ---
color.

The nine generators corresponding to an
infinitesimal 6D rotation involving $\w x^{\,5}$ or $\w x^{\,6}$ are denoted
by $\lambda_a$, where $a=0,\ldots,8$.
Eight of the nine HD generators $\lambda_1,\ldots,\lambda_8$
are colored and produce the strong interaction.
These generators correspond to rotations in planes with one HD coordinate
($\theta$ or $\phi$) and one 4D coordinate.
The ninth HD generator $\lambda_0$ is white and corresponds to the
known white elementary particles such as  the electron and photon.
This white generator corresponds to a rotation in the HD sphere of both HD 
coordinates $\theta$ and $\phi$, without a 4D coordinate. 
Fields associated with this generator are analogous
to $\w g_{55}=1$, which has two indices equal to five and has no HD
angular momentum.

We will construct our theory of the strong interactions based on our electroweak theory.
The generators of the HD part of the 6D rotation group are analogous to the spherical harmonics.
We will expand fields in terms of these nine generators, just like we expanded the vector $A_\m$ 
in terms of spherical harmonics. The reason for this is the same: it is because the
HD's are so small.

We expand $g_{\mu\nu}^{lm}$ and $A_\mu^{lm}$
in terms of the nine
generators $\lambda_{a}$ of the HD part of the 6D rotation group
\begin{eqnarray}
g_{\m\n}^{\,lm}&=&\sum_{a=0}^{8}g_{\m\n}^{\,lma}\la_{a},\label{eq:8.1}\\
A_\m^{lm}&=&\sum_{a=0}^{8}A_\m^{lma}\la_{a}.\label{eq:8.2}
\end{eqnarray}
The Latin indices $a,b,c$ range from zero to eight.
Henceforth we will observe the convention of summing over
these indices when they appear twice in a term. Up until now,
we have neglected the strong interactions. With these expansions, 
this is no longer possible. Previously defined fields
like the photon and graviton are now redefined---the photon and graviton
are no longer $A_\m^{00}$ and $g_{\m\n}^{00}$.  They are denoted now by
$A_\m^{000}$ and $g_{\m\n}^{000}$; the $W$ and $Z$ 
are now the $A_\m^{1m0}$.

In the 4D differential volume element, the
$\sqrt{g}$ refers to $g_{\m\n}^{000}$, which equals $\delta_{\m\n}$ in microscopic spacetime.
The $\sqrt{g}$ is not expanded in terms of HD rotation group
generators because $\delta_{\m\n}$ is constant and after an HD rotation,
the field's values are oriented the same way
in spacetime.  Therefore, it has no strong charge
with respect to an HD rotation.  Similarly,
in microscopic spacetime, the $A^\m_{lma}$ are written in terms
of the $A_\m^{lma}$ using the 4D contravariant
metric tensor $\delta^{\m\n}$ according to the formula
\begin{equation}
A_{lma}^\m=\delta^{\m\n} A_\n^{lma}.\label{eq:8.3}
\end{equation}
Thus, the only contravariant fields we will ever have to
consider are the $\delta^{\m\n}$ and the only generator we will ever have to
consider for the contravariant fields is $\la_{0}$.

Before we proceed to consider terms with factors expanded in terms
of HD rotation generators, we describe how to deal with the
generators in each term.
To eliminate the HD rotation generators from the terms in the Lagrangian
and make the terms scalars, we determine one generator for each of
the covariant and contravariant groups of factors in a term and
multiply the two generators together in a scalar or dot product. This
product is defined such that
\begin{equation}
\lambda_{a}\cdot\lambda_{b}=\delta_{ab}.\label{eq:B1}
\end{equation}
This is similar to integrating over the orthogonal kets representing 
the covariant and contravariant groups of factors.

Each factor in a term is expanded in terms of the nine
generators $\lambda_a$, eight of which are colored and one
is white. It remains to determine which type of product
exists between factors with these generators within each of the
covariant and contravariant groups.

There are three possible products for the $\lambda_a$ vectors
of the covariant factors:
scalar (dot product), vector (cross product), or a tensor product. 
In order to produce one
$\lambda_a$ for the covariant factors and yet
involve all of them, a cross product seems most appropriate
\begin{equation}
\lambda_a\times\lambda_b=[\lambda_a,\lambda_b]=
\sum_{c} c_{abc}\lambda_{c},\label{eq:B3}
\end{equation}
where the $c_{abc}$ are the structure constants of the HD
part of the 6D rotation group.
This is similar to the expression of products of kets in terms
of single kets using Clebsch-Gordan coefficients. 

Because the generators are not associative under cross product
multiplication, parentheses must be used to specify the
order of multiplications.  This would be the same as that
given in Sec.\,7.

According to the vector identity ${\bf A\cdot(B\times C)
=(A\times B)\cdot C}$, the dot product
may be interchanged with the last cross 
product in the term without affecting the value of the term.
This would be how to obtain the familiar form of mass and
Maxwell terms with two strongly interacting fields $A_\mu$
--- with the dot product between these factors as opposed
to between the covariant factors of $A_\mu$
and contravariant $\delta^{\m\n}$. The cross product that now exists
between the covariant and contravariant groups of factors drops out
because the contravariant group of factors has only the generator $\la_{0}$.
This leaves only a trivial ordinary multiplication between the covariant and
contravariant groups.

Expanding Terms~(\ref{eq:5.1}) first in
terms of spherical harmonics and then in terms of HD rotation generators,
we arrive at the terms
\q
-\tfrac14\,
\overline F^{\,\m\n}_{00a}F_{\m\n}^{00a}
+\tfrac{1}{4\sqrt3}\,\overline F^{\,\m\n}_{11a}F_{\m\n}^{11a}
+\tfrac{1}{4\sqrt3}\,
\overline F^{\,\m\n}_{10a}F_{\m\n}^{10a}
+\tfrac{1}{4\sqrt3}\,
\overline F^{\,\m\n}_{1,-1a}F_{\m\n}^{1,-1a},\label{eq:8.4}
\qq
where $F_{\m\n}^{lma}=\partial_\m A_\n^{lma}
                         -\partial_\n A_\m^{lma}$.

The term with one factor of $A$, with only this factor expanded
in terms of colored rotation generators, is zero due to the orthogonality
of the generators.  (With only one factor of colored $\lambda_a$, there
is no way to make the term a scalar.)

Doubly expanding the
factors of $A$ in Terms~(\ref{eq:5.3})--(\ref{eq:5.4a}), we obtain
\begin{eqnarray}
+\ \frac{9+4\sqrt3\,}{12c^2 T^2}\,
&A^\m_{\,11a}A_\n^{11a}\label{eq:8.5a}\\
       +\ \frac{9+4\sqrt3\,}{12c^2T^2}\,
&A_\m^{\,1,-1a}A_\n^{1,-1a}\label{eq:8.5b}\\
        +\ \frac{\sqrt{30}\,}{3c^2T^2}\,
&A_\m^{\,10a}A_\n^{10a},\label{eq:8.5c}
\end{eqnarray}  
which are  mass terms for the $A_\m^{1ma}$.

The Lagrangian for 3D time should
contain terms with anywhere from two to six strongly interacting
vectors or gluons.  However, strong interactions with diagrams with
five or six external gluon lines, corresponding to terms with five or six
gluons, are not observed. 3D time provides the explanation
for this. If one examines the definition of the curvature scalar
described in Sec.\,4, one finds that the only way to
obtain terms with five or six factors of $A_\mu$ is through the
presence of $\widehat g^{\,55}=1+g^{\mu\nu}A_\mu A_\nu$. Now
since the strong interaction cross product between the two factors
of $A$ here is purely antisymmetric, while the multiplication of these two
factors by $g^{\mu\nu}$ symmetrizes them, the resulting terms
with five or six gluons are zero.
\pagebreak[2]

{\samepage
\begin{center}
{\bf 11. Quarks, confinement, asymptotic freedom and chiral symmetry breaking}
\end{center}

\nopagebreak
The main aspects of strong interaction phenomenology are quarks,
quark confinement, asymptotic freedom and chiral symmetry breaking.
These are not hard to achieve in 3D time.
The existence of quarks can be explained by expanding the photon
in terms of HD rotation generators and spinors. In final form,
for example
\q
A_\m^{00}|00\rangle=-\tfrac12\s_\m^{B\dot V}
\chi_B^p\varpi_{\dot V}^q\lambda_{pq}\tfrac{1}{\sqrt2}\k\kk
+\tfrac12\s_\m^{B\dot V}\zeta_B^p\iota_{\dot V}^q
\lambda_{pq}\tfrac{1}{\sqrt2}\kk\k.\label{eq:9.1}
\qq}
We have neglected to consider the terms corresponding to the
one white generator in the expansion for
$A_\m^{00}$.  This is the photon. 
The index for the eight colored generators of the HD rotation group
has been rewritten as $pq$ where $p$ and $q$ take the values
r,g,b, which stand for red, green and blue.  A sum over $p$ and
$q$ is implied.  The spinors $\chi_B^q$ and $\iota_{\dot V}^p$
are associated with the ket $\k$, while $\varpi_{\dot V}^p$ and
$\zeta_B^q$ have $\kk$ as their HD angular
momentum representation. Transforming the spinors according to 
Sec.\,8, we find the up quark $u^p$ corresponds to $\iota_{\dot V}^p$
while the down quark $d^p$ corresponds to $\varpi_{\dot V}^p$.

By expanding $A_\m^{00}$ in terms of rotation generators and spinors
with $l>\frac12$, quarks in other families may be produced. The
strange and charmed quarks are associated with the states
$\left |\tfrac32,-\tfrac12\right >$ and
$\left|\frac32\frac12\right>$, respectively.  The top (bottom)
quark is in the state $\left|\frac52\frac12\right>$
$\left(\left|\frac52,-\frac12\right>\right)$. According to
3D time, there are an infinite number of quarks, each 
corresponding to a state $|lm\rangle$, where $l$ and $m$ are
half-odd integral. Thus, expanding $A_\mu$ in terms of spherical
harmonics, rotation generators and spinors produces multiple
generations of quarks. It is not unlikely that the number of
generations appears to be limited by the same reason given in Part~II
for leptons.

Terms for the interactions of quarks with
gluons may be derived from the gluons' mass terms.
This occurs by substituting a quark-antiquark pair for one gluon in the gluon mass terms.
It is highly unlikely that quarks have charges exactly one-third
that of the electron because the value of a particle's charge
will be shown in Part~II to result from quantum corrections
and these are notoriously uneven in value.
There does not appear to be any way of achieving one-third integral
charges from the methods used in Part~II.
Thus, we will subscribe to the original theory of the strong interactions,
the Han-Nambu model [59,60],
in which quarks have integral charges. This is possible because our theory
does not have an exact strong interaction symmetry of the Lagrangian.
Quarks of different colors have different charges. The symmetry is broken.
Instead of quarks having color, they are non-singlet $SU(3)'$ states in
the Han-Nambu model. Thus, when I use the word `color,' I am referring
to this property and not color as it is usually used.
 
Terms describing free quarks (the Dirac term), quark masses and the
electromagnetic interactions of single quarks are zero.  This is because the
colored generators of the quarks are orthogonal to the white generator
of $\delta^{\m\n}$ or $A_\m^{000}$ (the only other fields in
the terms) and their dot or scalar product is zero.  With only one set of
colored generators there is no way to make the term a scalar.

The SM assumes the Dirac or free-quark term
in its Lagrangian, thereby postulating free quarks.
It then goes to great lengths to eliminate the same
free quarks. Why not just leave out the free-quark term?
In 3D time,  the free-quark term is zero.  This means there are no
free quarks.  How can one have free quarks
without a free quark term in the Lagrangian to describe them?
The only way a quark can exist is if it is interacting with a gluon.
Certainly this method of quark confinement is much simpler than the
charge-antiscreening-due-to-vacuum-polarization method,
which requires the use of supercomputers for its calculations and still does not work completely.
My method requires virtually no calculations, is far simpler and is more effective.
And it is just another example of determining the correct Lagrangian for each field of interest.
Sometimes it is not what you put in a Lagrangian, but what you leave out.

Although there are no mass terms for quarks in $\widehat R$,
masses for quarks can be generated by their color, in much the same
way that the classical radius equation can be used to generate a
mass for the electron from its charge. Thus, we would have only strong
interaction and mass terms for quarks. New terms for quarks, however, can
be generated by expanding the white vectors $A_\mu^{lm0}|lm\rangle$
in terms of spinors that are white combinations of quarks. In this case,
the terms that applied to previously considered spinors now apply to white
combinations of quarks. 
In this way, we arrive at free-field and photon interaction terms
for white quark combinations known as baryons. Because the
free-quark terms vanish but the free-baryon terms do not, quarks
can only be found in white combinations, of which baryons is one class.

The theory of 3D time's strong interaction sector may be tested 
by detecting free gluons.  These are allowed in 3D time because the
free-field (Maxwell) terms for gluons do not vanish.
Free gluons are forbidden by confinement in the SM.

The chiral symmetry breaking of the strong interactions
is caused by a nonzero pion mass.
According to the classical radius equation presented in Part~II,
a particle that has charge must have mass.  The pion has a mass
because the quarks and gluons of which it is composed have
masses due to their strong interaction charges or color.
Therefore, 3D time explains chiral symmetry breaking.

We now describe an effective potential for confinement in 3D time.
Each of the gluons with which the quarks interact
must have a mass because it has color,
which would, like electric charge,
yield a mass by the classical radius equation in Part~II.
Because free quarks are forbidden by the vanishing of the
free-quark term they must always be interacting.  Now, in order
for two quarks to remain interacting, they must be within range
of the strong interaction.  Therefore, they cannot be separated by
a distance greater than the Compton wavelength
$\lambda=h/mc$ of a gluon.
The mathematical expression for this is
an infinite square-well potential with radius equal to the 
Compton wavelength of a gluon surrounding each quark.
Inside this infinite square well, the quarks are free to
move about because the gluons that influence
the quarks have masses due to their color, rendering the
interaction weak.  Thus, aside from the infinite square-well
confining potential, the strong interaction is really weak.
This finishes the explanation of quark confinement and asymptotic freedom.

\clearpage
\begin{center}
{\bf Part II. The Quantum Theory}
\end{center}

\begin{center}
{\bf 12. Quantum field theory in 3D time}
\end{center}

Now that we have the classical theory of 3D time, we must now calculate
the quantum corrections to it. This results in different values
for some of the coefficients (masses and charges) of its terms. In particular, we will derive
the values of ten previously unexplained parameters in the Standard Model (SM).
Additionally, masses of seven of 3D time's predicted particles will be determined.

Quantum mechanics is not philosophical in origin.
It results in part from the properties of waves. For example, it is
the fact that one cannot simultaneously determine the position and
momentum of a wave that leads to the Uncertainty Principle.
If a wave  is narrow in momentum space, it must be spread out 
or have many possible values in configuration space.
If it occupies a narrow location in configuration space 
it must be spread out or have many possible values in
momentum space. This is simply a property of waves.
The waves follow from wave equations, which are field equations.
The field equations originate from the Lagrangian.
The Lagrangian is the curvature scalar of 3D time.
The curvature scalar is the Lagrangian for general relativity.
Therefore, the Uncertainty Principle is derived from general relativity.
Proceeding in this manner, it seems to me likely that one can derive 
all aspects of quantum mechanics from the curvature scalar or in other words, general relativity.
Until this is done we will merely assume quantum mechanics comes from
general relativity and use it freely. The same applies to quantum field theory,
which is the logical extension of quantum mechanics to fields.

Particles in 3D time are extended
with sizes of the order of the Planck length. Particles were thought to have to be 
pointlike in order to avoid violations of either special relativity or causality, but 
string theory has shown that extended particles can, in fact, be viable.
With the properties of elementary particles in 3D time,
such as a particle's nonzero size and a tiny mass for the photon,
one can replace QED by a simplified theory that does not have regularization,
renormalization, and gauge invariance. I will begin to do this here
by using one-loop quantum correction equations from present-day QED.

Renormalization is the method for handling infinities
when particles are pointlike but if the particles are extended,
the infinities never develop. Instead, there are cutoffs.
The cutoffs cause quantum corrections to be finite,
which allows the calculation of actual quantities from finite bare quantities.
Therefore, extended particles eliminate the need for renormalization
and allow the calculation of previously unexplained parameters.
For example, consider the equation~[61]
\begin{equation}
m=m_0+\delta m
\end{equation}
from QED.  Here $m$ is the mass of the electron,
$m_0$ its bare mass and $\delta m$ the quantum mass correction.
In a renormalizable model, both $m_0$ and $\delta m$ are infinite
and cancel and $m$ must be inserted by hand.  Thus, $m$ here is
completely unexplained.  However in 3D time, both $m_0$
and $\delta m$ are finite and precisely determined,
allowing a determination of $m$ to be made.
In this way, one can calculate masses.
Similarly, the equation in QED for the quantum charge correction
allows one to calculate coupling constants.

We will now show how to determine $m$.
The equation for the one-loop electron
quantum mass correction [61] for electromagnetism is
\begin{equation}
m=m_0+\delta m,\label{eq:13.1}
\end{equation}
where
\begin{equation}
\delta m=\frac{3\alpha}{4\pi}m\left(\ln\frac{\Lambda^2}{m^2}
+\frac12\right).\label{eq:13.2}
\end{equation}
Here $\alpha$ is the fine structure constant. 
The mass $m_0$ for the electron is apparently its classical free mass.
This is the mass when the charge of the
electron is zero and without a quantum mass correction.
The mass $\Lambda$ is the classical
interacting mass for the electron.  
This is the mass of the electron when it has charge, but before the
quantum mass correction is made.
The mass $m$ is the quantum-corrected interacting mass.
This is the mass of the electron with its charge and after the quantum mass
correction has been taken. It is the actual mass of the electron.

In 3D time, the mass $m_0$ is zero [from Part~I].  This is
because, unlike vectors and tensors, there are no mass terms in 
the 6D curvature scalar $\w R$
for spinors.  In addition, because $m_0$  is calculated with
its charge equal to zero, the electron does
not acquire a mass from its charge.  Producing
mass from charge is described in the next section.

We have
\begin{equation}
m_0=0.\label{eq:13.3}
\end{equation}

If one distrusts the math in this document, one is encouraged to do the
following calculation. It is both extremely simple and extremely powerful.
The hierarchy problem can be solved in five minutes. All one need do is
combine Eqs.\,(\ref{eq:13.1}), (\ref{eq:13.2}) and~(\ref{eq:13.3})
and solve for $m$. The result is the mass formula
\begin{equation}
m=\exp\left(\frac14-\frac{2\pi}{3\alpha}
    \right)\Lambda.\label{eq:13.4}
\end{equation}
If one substitutes a Planck mass for $\Lambda$, an elementary particle mass results for $m$.
QED.

This solution to the hierarchy problem is the most important equation 
in the quantum part of this document. It converts 
Planck masses $\Lambda$ to elementary particle masses $m$, without fine tuning.
This allows predictions at the Planck energy to be tested at low energy.
It renders observable the HD coordinate dependence in a Kaluza-Klein
theory by converting the masses of the higher order harmonics from
Planck masses to elementary particle masses. The massive harmonics
become the elementary particles. We have simply started with the right
postulates and are trying to determine the quantum corrections 
to our classical theory. This equation
follows directly from the classic quantum field theory textbook, 
Itzykson and Zuber [61]. Talk about being hidden in plain sight!
The answer to the hierarchy problem was right under our noses the whole time!

The problem of deriving the mass of the electron on purely
theoretical grounds now translates into finding the proper values
for $\Lambda$ and $\alpha$. The search for the values of these
parameters for each elementary particle occupies the bulk of the
remainder of this document. If one substitutes the Planck mass for
$\Lambda$ and sets $\alpha$ to 1/137, the mass $m$ obtained is
much smaller than the electron mass.  This problem is opposite the
usual one, where the mass of the elementary particle is too large. 
The question arises,
``What will $\alpha$ have to be to convert a Planck mass to an
ordinary mass?"  If Eq.\,(\ref{eq:13.4}) is solved for $\alpha$,
with $\Lambda$ the Planck mass and $m$ the mass of the electron,
one finds $\alpha$ must be larger than 1/137.  
Perhaps $\alpha$ in Eq.\,(\ref{eq:13.4}) is the bare fine
structure constant. The bare fine structure constant should be
larger than 1/137.

As will be shown, the fine structure constant $\alpha$ starts
out at 1, is then reduced to 1/19 by charge screening due to
vacuum polarization and finally reduced to 1/137 by the
vertex correction. Because charge screening due to vacuum
polarization is allowed by the Feynman diagram for the mass of
the electron, while the vertex correction, which has an external
photon line, is not, the value for $\alpha$ in Eq.\,(\ref{eq:13.4})
is 1/19. This means the ordinary mass scale is actually the quantum-corrected Planck mass.

Section~13 derives an expression for $\Lambda$.
This expression contains the photon's radius, which is
calculated in Sec.\,14.
Section~15 determines the value 1/19 for $\alpha$ in
the quantum mass correction equation.
Section~16 explains the masses of the electron, muon, tau and  photon
and the value for the fine structure constant. Masses for the
neutrinos are deduced in Sec.\,17.
New fermions are described in Sec.\,18. The reason why only the observed 
elementary particle interactions take place is explained in Sec.\,19.
Section~20 derives the masses of the $W^\pm$ and $Z^0$ from scratch
(instead of from the weak interaction coupling constant $G_W$ as is presently
done in the Standard Model).
One of the main predictions of the theory --- new intermediate vector bosons
with precisely determined masses --- is given in Sec.\,21.
The so-called Higgs boson is discussed in Sec.\,22,
which is where 3D time's predictions are summarized.
Conclusions are presented in Sec.\,23.
\pagebreak[2]

{\samepage
\begin{center}
{\bf 13. The classical photon mass $\Lambda$}
\end{center}

\nopagebreak
The electron does not have a mass term in the 
6D curvature scalar $\w R$.  However, the electron
acquires a mass through its electric field.
An electric field contains energy and since the electron
always carries this field with it, this energy may be
considered to be a rest mass.}  Equivalently, the source of the
electron's mass may be ascribed to its charge.  It requires energy
to keep together an assemblage of electric charges of the same sign,
which tend to repel each other. This is represented by the
equation for the electron's classical radius
\begin{equation}
mc^2=\frac{e^2}{r_c},\label{eq:14.1}
\end{equation}
where $e$ is the electron's charge and $r_c$ its classical radius,
which is of the order of the size of its charge distribution.  Equation~(\ref{eq:14.1})
shows why a particle that has charge must have mass.
This origin of mass is one of the two sources of mass in 3D time.  
The other is explicit mass terms in the 6D curvature scalar.
One reason the classical radius origin of mass is not more accepted 
is that the classical radius equation concept as it stands is flawed; 
the situation is fixed in Sec.\,14.

The remainder of this section is essentially a derivation of the
classical radius equation for the case of the photon. 
We choose the photon because $\Lambda$ in QED is defined [61] to be
the large mass of a ficticious massive photon. I contend, however,
that $\Lambda$ is the Planck mass of the real photon, that is,
its mass before the quantum mass correction. We will have to calculate
$\Lambda$ from the photon's charge because there is no mass term
for the photon in $\w R$. This section also
demonstrates how to construct an elementary particle
by determining its distribution of charge.

The photon must have some type of charge 
because anything that interacts with the photon can be said to have 
charge and one photon interacts with another during
photon-photon scattering [61]. The Feynman diagram for photon-photon scattering has
four external photon lines, four internal electron lines, four vertices and one loop.
In this interaction, photons do not interact
with each other directly, but indirectly through electrons. Thus the actual charge of
each photon is zero. Nevertheless, it is instructive to assign to each of the two photons involved
an effective charge as if the two photons had interacted directly.
The four vertices in the diagram for
this interaction imply that the coupling constant
is proportional to $e^4$, where $e$ is the charge on an electron.
I claim the effective charge $q_\gamma$ for each of the two photons in this interaction is
\q
q_\gamma=\frac{e^2}{4\pi\sqrt{\hbar c}}.\label{eq:14.1a}
\qq

In analogy with the electron, the classical interacting mass for the photon 
will be equal to its electrostatic energy $H$. Equation~(\ref{eq:14.1})
for the classical radius of the electron is derived from classical
electrodynamics but is valid for quantum electrodynamics as well.
Therefore, we will derive the classical interacting mass of the photon from
classical electrodynamics~[62]. We have
\begin{equation}
\Lambda c^2=H,\label{eq:14.2}
\end{equation}
where the electrostatic energy is derived from the electric field
${\bf E}$
\begin{equation}
H=\frac{1}{8\pi}\int |{\bf E}|^2\,d^3x.\label{eq:14.2a}
\end{equation}
The electric field is obtained from the potential $\Phi({\bf x})$
\begin{equation}
{\bf E}=-\overrightarrow\nabla\Phi({\bf x}).\label{eq:14.2b}
\end{equation}
Finally, the potential is derived from the charge density $\rho({\bf x})$
\begin{equation}
\Phi({\bf x})=\int\frac{\rho({\bf x^{\,\prime}})}{|{\bf x}-{\bf x}^{\,\prime}|}
\,d^3x^{\,\prime}.\label{eq:14.2c}
\end{equation}
A prime on a coordinate means that is the coordinate for a point in the
charge density.
Because ${\bf x}^{\,\prime}$ is the variable of integration and ${\bf x}$
is the observation point, it is useful to separate the variables in
$1/|{\bf x-x}^{\,\prime}|$
\begin{equation}
\frac{1}{|{\bf x-x}^{\,\prime}|}=4\pi\sum^\infty_{l=0}\sum^l_{m=-l}
\frac{1}{(2l+1)}\,\frac{r^{\prime l}}{r^{l+1}}Y^*_{lm}(\theta^{\,\prime},\phi^{\,\prime})
Y_{lm}(\theta,\phi),\label{eq:14.2d}
\end{equation}
where $r=|{\bf x}|$ and the $Y_{lm}(\theta,\phi)$ are the spherical
harmonics. The next step is to determine the charge density
$\rho({\bf x}^{\,\prime})$.

The charge is located, of course, where the
particle is located. The particle is located in ordinary 
three-dimensional space (3DS). However, one cannot localize the
photon field to a distance smaller than $2r_\gamma$, where $r_\gamma$
is the radius of the photon, along its direction of motion 
because this would mean that it would be localized
on the HD sphere. This localization on the HD sphere
is not possible because the photon's higher-dimensional wave function,
which is the spherical harmonic $Y_{00}(\theta,\phi)=1/\sqrt{4\pi}$,
is by definition spread out uniformly over the HD sphere.
The coordinate $\theta$ of the sphere projects
onto the photon's 4D world line, which is the $z$-axis
of the embedding space of the HD sphere.
The photon's quantum
mechanical spin causes the length $2r_\gamma$ to form a
sphere of radius $r_\gamma$ in 3DS. The photon is therefore
a line segment of a certain type of charge
of length $2r_\gamma$ spinning about its
midpoint.

The charge density of the photon will be proportional to the magnitude squared
of its wave function. This is because the charge is located where the particle 
is located and the probability of finding the photon at a given point is equal 
to the magnitude squared of its wave function. We have
\q
\rho({\bf x}^{\,\prime})=q_\gamma{|\psi({\bf x}^{\,\prime})|}^2.\label{eq:14.3a}
\qq
The wave function
$\psi({\bf x}^{\,\prime})$ of a photon as localized as possible in 3DS is equal 
to its radial wave function $R(r^{\,\prime})$ times its spin wave function
$S(\theta^{\,\prime},\phi^{\,\prime})$
\begin{equation}
\psi({\bf x}^{\,\prime})=R(r^{\,\prime})S(\theta^{\,\prime},\phi^{\,\prime}).\label{eq:14.3}
\end{equation}
The radial wave function is given by the projection of the HD
wave function onto the 4D world line parameterized by the coordinate
$z$. As with any projection of an object, this projection
will be proportional to the size of the original object,
in this case the HD wave function. 
Therefore, the radial wave function $R(r^{\,\prime})$ in 3DS is proportional to the
higher-dimensional wave function $Y_{00}$, which is constant.
This result applies to any $Y_{lm}$ --- 
the projection of $Y_{lm}$ onto the $z$-axis
is proportional to $Y_{lm}$. 

After normalization, the radial function is given by a new constant $c$
\begin{eqnarray}
R(r^{\,\prime})&=&c,\qquad r_\gamma \geq\ r^{\,\prime}\geq\ 0\label{eq:14.5a'}\\       
R(r^{\,\prime})&=&0, \qquad \infty>\ r^{\,\prime}>r_\gamma.\label{eq:14.5b'}
\end{eqnarray}
The spin wave function is given by
\begin{equation}
\frac{1}{\sqrt3}\left(
\begin{tabular}{c}
1\\
0\\
1
\end{tabular}
\right)
\to \tfrac{1}{\sqrt3}Y_{11}(\theta^{\,\prime},\phi^{\,\prime})
+\tfrac{1}{\sqrt3}Y_{1,-1}(\theta^{\,\prime},\phi^{\,\prime})
=S(\theta^{\,\prime},\phi^{\,\prime}),
\label{eq:14.6}
\end{equation}
where we have omitted the longitudinal polarization
state $Y_{10}(\theta^{\,\prime},\phi^{\,\prime})$ of the photon.
The constant $c$ is determined by the condition
\begin{equation}
\int \rho({\bf x^{\,\prime}})\,d^3 x^{\,\prime}=q_\gamma.\label{eq:14.7}
\end{equation}
We find
\begin{eqnarray}
\rho({\bf x}^{\,\prime})&=&\frac{9q_\gamma}{8\pi r_\gamma^{\ 3}}\,
\sin^2\!\!\theta^{\,\prime},\qquad r_\gamma\geq\ r^{\,\prime}\geq\ 0\label{eq:14.8a}\\
\rho({\bf x^{\,\prime}})&=&0,\qquad\qquad\qquad \infty>\ r^{\,\prime}>r_\gamma.\label{eq:14.8b}
\end{eqnarray}
Substituting this into Eq.\,(\ref{eq:14.2c}),
it follows from Eqs.\,(\ref{eq:14.2a})--(\ref{eq:14.2d}) that
\begin{equation}
H=\frac{q_\gamma^{\ 2}}{2r_\gamma}.\label{eq:14.9}
\end{equation}
This is the monopole contribution in Eq.\,(\ref{eq:14.2c}). The higher order
terms are negligible. Therefore, the dependence of $H$ upon the
angular and radial dependence of $\rho({\bf x}^{\,\prime})$ is negligible.
The energy behaves as if the charge were uniformly distributed
throughout a sphere of radius $r_\gamma$.

Equations~(\ref{eq:14.9}) and~(\ref{eq:14.2}) yield
for the classical mass of the photon
\begin{equation}
\Lambda=\frac{q_\gamma^{\ 2}}{2c^2 r_\gamma}
         =\frac{\beta\hbar}{2c\,r_\gamma},\label{eq:14.10}
\end{equation}
where 
\q
\beta=\frac{q_\gamma^{\ 2}}{\hbar c}=\frac{e^4}{(4\pi)^2\hbar^2 c^2}\label{eq:14.10a}
\qq
is the fine structure constant for the photon.
Equation (\ref{eq:14.10}) is the classical radius equation.
The distance appearing in the classical radius equation is
that between the two most widely separated points of charge
in the distribution.  For a spherical distribution it is therefore
the diameter.  This is why a factor of two multiplying the radius appears in our equation.  
This factor does not appear in the usual definition of the 
classical radius equation, which is only an
order-of-magnitude estimate.

The effective charge on a photon is proportional to the square of the charge on an electron.
It turns out that the fine structure constant for the photon $\beta$ is always equal 
to the square of the fine structure constant for the electron. 
This holds true at every point in the calculation for $\alpha$, the fine structure constant
for the electron, done later. Using the relation
\q
\beta=\alpha^2,\label{eq:14.10b}
\qq 
we have
\begin{equation}
\Lambda=\frac{\alpha^2\hbar}{2c\,r_\gamma}.\label{eq:14.10'}
\end{equation}
It remains to determine the radius $r_\gamma$ of the photon.
\pagebreak[2]

{\samepage
\begin{center}
{\bf 14. The radius of the photon}
\end{center}

\nopagebreak
In order to eliminate the infrared divergences in QED, the photon must
have a small mass.
Therefore, we treat it in what follows as a massive particle, like
the electron. Among other things, this means the photon may be
brought to rest. If a particle is at rest and located at the
origin of ordinary three-dimensional space (3DS), then its 
4D world line is the time axis.
However, particles are never really at rest
because they are spinning. Setting the photon's spin angular
momentum equal to $\Lambda v \langle r \rangle$, we have
\begin{equation}
\left[s(s+1)\right]^{1/2}\hbar=\Lambda v \langle r \rangle.\label{eq:15.1}
\end{equation}}

A simple way of determining the radius at which the particle
rotates is to place its entire mass at a particular average radius
$\langle r \rangle$. 
The first step in determining $\langle r \rangle$ is to determine $R(r)$ because
\q
\left<r\right>=\int_0^{r_\gamma} R^*(r)\,rR(r)\,dr.\label{eq:15.1a}
\qq
The quantity
$R(r)$ will be the projection of the photon's higher-dimensional wave function $Y_{00}$,
which is constant, onto its 4D world line. 
Any projection is proportional to the size of the original
object so $R(r)$ must be proportional to a constant.
The second requirement on $R(r)$ is that it be normalized to unity.
The normalization criterion is $\int_0^{r_\gamma} R(r)^* R(r) dr = 1$. 
We find
\q
R(r) = \frac{1}{\sqrt{r_\gamma}}.\label{eq:15.1b}
\qq
This radial wave function for the photon is constant and satisfies 
the normalization criterion. 
Substituting Eq.\,(\ref{eq:15.1b}) into Eq.\,(\ref{eq:15.1a}), we find
\q
\langle r \rangle = \frac{1}{2}\,r_\gamma.\label{eq:15.1c}
\qq

Setting $s=1$ in Eq.\,(\ref{eq:15.1}), we obtain
\begin{equation}
\Lambda v \tfrac 12\,r_\gamma=\sqrt 2\,\hbar.\label{eq:15.1'}
\end{equation}
Substituting Eq.\,(\ref{eq:14.10}) into Eq.\,(\ref{eq:15.1'}), we have
\begin{equation}
v=\frac{4\sqrt2\, c}{\beta}.\label{eq:15.2}
\end{equation}

Note that if $\beta=\alpha^2=(1/137)^2$, the velocity
of spin for the photon is much greater than the speed of light.
This is the main objection to the concept of the classical radius
equation.  There is a way around this problem, however.  The trick
is to use the bare fine structure constant
$\beta_{0}=\alpha_0^{\ 2}$, which will later be determined
to be unity, in the denominator of Eq.\,(\ref{eq:15.2}) instead of the
actual fine structure constant.
As we will show, this larger denominator will bring the
velocity of spin for the photon to something below the speed of
light.  The justification for substituting the bare fine structure
constant for the actual one is that only the bare photon is
spinning.  Its surrounding outer shell of virtual particles caused by
vacuum polarization and the vertex correction
does not spin along with it.

We will define the bare fine structure constant $\beta_{0}$
for the photon to be
$
\beta_{0}=\alpha_0^{\,2}
$
while the actual fine structure constant $\beta$ is
\q
\beta=4\pi\alpha^{\ \ 2}_{Ve},\label{eq:15.2ab}
\qq
where $\alpha_{Ve}$ is the fine structure constant for the electron after all
quantum corrections have been taken, except those involving the muon or tau. 
The addition of the fine structure constants involving vacuum polarization 
with the muon and tau to that of the electron only, coincidentally add up to an 
additional factor of  $\sqrt{4\pi}$ multiplying each factor of $\alpha_{Ve}$.
There are two such factors in Eq.\,(\ref{eq:15.2ab}), producing a factor
of $4\pi$. Vacuum polarization involving the muon and tau is discussed in Sec.\,16.
This introduces a factor of $4\pi$ into the
denominator of Eq.\,(\ref{eq:15.2}) when the actual fine structure constant
is converted to the bare one with vacuum polarization involving the muon and tau.
Thus
\begin{equation}
v=\frac{\sqrt2\,c}{\pi\alpha_0^{\ 2}}.\label{eq:15.2'}
\end{equation}

We now show why $\alpha_0=1$.
We start with the z-component of angular momentum operator
$L_z$ operating on the neutrino's
and electron's wave functions $\psi_\nu$ and $\psi_e$
\begin{equation}
L_z
\left(
\begin{array}{c}
\psi_\nu\\
\psi_e
\end{array}
\right)=\frac{\hbar}{2}
\left(
\begin{array}{cc}
1&0\\
0&-1
\end{array}
\right)
\left(
\begin{array}{c}
\psi_\nu\\
\psi_e
\end{array}
\right)
\label{eq:15.3a}
\end{equation}

This shows the electron has z-component of angular momentum $-\frac{1}{2}\hbar=m\hbar$
in the three-dimensional embedding space of the HD sphere.
This angular momentum is more commonly known as isospin.
Here $m$ does not represent a mass but rather the quantum number
for the z-component of isospin. The electron's charge quantum number,
which is integral and $m$ are related by elementary
quantum mechanics according to the relation [63]
\begin{equation}
m=c+\mbox{int.},\label{eq:15.7}
\end{equation}
where $c$ is a constant for the multiplet and int.\ is an integer.
According to Ref.\,[63],
$c=\frac 12$ for spinors. 
Equation (\ref{eq:15.7}) is the Gell-Mann relation
$Q=I_3+\frac12 y$, with the following identifications:  charge
$Q=\mbox{int.}$, the z-component of weak isospin $I_3=m$ and one-half
the weak hypercharge $\frac12 y=-c$. Therefore, the origin of the Gell-Mann
relation is explained by 3D time.

If
\q
Q=m-1/2,
\qq
which is a restatement of the Gell-Mann relation, and $m=-1/2$
corresponds to an angular momentum of $m\hbar$, then $Q=-1$
must correspond to an angular momentum of $Q\hbar$. We have
\q
L_Q=Q\hbar, \label{eq:15.3aa}
\qq
where $L_Q$ is the angular momentum due to the electron's charge. 
The equation defining the magnitude of this angular momentum is 
\q
L_Q=\left|(\oo r\times \oo p)\right|_Q=rp_Q\sin\theta_Q, \label{eq:15.3b}
\qq
where $r$ is the radius of motion, $p_Q$ is the linear momentum
and $\theta_Q$ is the angle between $\oo r$ and $\oo p$. 
The subscript $Q$ means the quantity pertains only to the
angular momentum of the electron's charge.
Combining Eqs.\,(\ref{eq:15.3aa}) and (\ref{eq:15.3b}), we have
\q
L_Q=cTmv_Q\sin\theta_Q=Q\hbar, \label{eq:15.3c}
\qq
where $cT$ is the radius of the HD sphere and is the radius of motion and
$mv_Q$ is the magnitude of the linear momentum. 
The quantity $cT$ is the Planck length.
We have 
\q
cT = \sqrt{\hbar G/c^3}. \label{eq:15.3e}
\qq
Combining Eqs.\,(\ref{eq:15.3e}) and (\ref{eq:15.3c}) we have
\q
mv_Q\sin\theta_Q=Q Mc=p^{\,\prime}, \label{eq:15.3f}
\qq
where 
\q
M=\sqrt{\hbar c/G} \label{eq:15.3ff}
\qq
is the Planck mass. The energy for this type of motion is of the form
$p^2/(2m)$. In our case it follows from Eq.\,(\ref{eq:15.3f}) and is
\q
E=\frac{p^{\,\prime\, 2}}{2m}=\frac{Q^2Mc^2}{2}. \label{eq:15.3g}
\qq
From the classical radius equation we have
\q
E=\frac{e_0^{\,2}}{2cT}, \label{eq:15.3h}
\qq
where $e_0$ is the electron's bare charge and $2cT$ is the diameter
of the HD sphere. 

The energy of motion of the electron on the HD sphere
that causes its charge is equal to its electrostatic self-energy. Therefore,
equating the two energies in Eqs.\,(\ref{eq:15.3g}) and (\ref{eq:15.3h})
and using Eqs.\,(\ref{eq:15.3e}) and (\ref{eq:15.3ff}), we have
\q
e_0^{\,2}=Q^2\hbar c, \label{eq:15.3i}
\qq
The definition of the bare fine structure constant $\alpha_0$ is
\q
\alpha_0=\frac{e_0^{\,2}}{\hbar c}. \label{eq:15.3j}
\qq
Combining Eqs.\,(\ref{eq:15.3i}) and (\ref{eq:15.3j}) produces the result
\q
\alpha_0=Q^2.\label{eq:15.3k}
\qq
The electron has $Q=-1$, so for it, $\alpha_0=1$ as was to be shown.
A similar derivation may be made involving the total angular momentum 
operator squared $L^2$ instead of $L_Q$. This leads to the result $\alpha_0=l(l+1)$.
This is the coupling constant for the weak interactions as they couple to isospin.

Substituting this into Eq.\,(\ref{eq:15.2'}), we have
\begin{equation}
v=\frac{\sqrt2}{\pi}\,c.\label{eq:15.2''}
\end{equation}
Because $\sqrt2$ is less than $\pi$, the velocity of spin is less
than the speed of light.  This is the velocity of spin of a point
midway out from the center of a spinning photon. The velocity of a
point on its surface is twice as great but is still less
than the speed of light because $2\sqrt2$ is less than $\pi$.
As will be implied shortly, this velocity of spin applies only
to an image of the photon, not the actual photon itself.
The velocity of spin of the actual photon is considerably less.

The size of the charge distribution in 3DS is
given by its projection from the 4D world line onto a spatial
axis in 3DS. Therefore $r_\gamma=cT \sin\psi$, where $\psi$
is the angle between the 4D world line and the ordinary $t$-axis.
Thus the radius for the photon is less than $cT$. However,
we now convert the photon's charge density of a solid sphere in 3DS into the
surface charge density of a hollow sphere. The two are equivalent 
from our discussion of the classical radius equation. The energy behaves as if 
the internal distribution of charge does not matter. Now, the HD sphere
forms a boundary in 3DS because the higher dimensions overlap 3DS. A spherical
surface charge density inside the HD sphere plus a boundary at the
HD sphere is equivalent to a spherical surface charge density outside the HD sphere.  
This follows from the method of images in classical electrodynamics.  
Therefore, we have
\q
r_\gamma^{\,\prime}=\frac{c^2T^2}{r_\gamma},
\qq
where the effective radius of the photon is $r_\gamma^{\,\prime}$.
Substituting $r_\gamma=cT\sin\psi$ into this equation,
we find
\q
r_\gamma^{\,\prime}=\frac{cT}{\sin\psi}.
\qq
Henceforth we will drop the prime from $r_\gamma^{\,\prime}$
\q
r_\gamma=cT/\sin\psi\label{eq:15.4}
\qq
and when the radius of a particle is referred to, we will mean 
its effective radius.

From the photon's effective velocity of spin determined above, one may calculate
the angle $\psi$ from the following equation
\begin{equation}
\tan\!\psi=\frac{\Delta x}{c\Delta t}=\frac{v}{c}=
             \frac{\sqrt2}{\pi},\label{eq:15.5}
\end{equation}
where $x$ is the spatial path in which the photon moves as it spins
and the time axis is labeled $ct$.
Therefore,
\begin{equation}
r_\gamma=2.4361cT.\label{eq:15.4'}
\end{equation}

Combining Eqs.\,(\ref{eq:15.6}), (\ref{eq:15.4'}) and (\ref{eq:14.10'})
we obtain the value for the classical mass $\Lambda$ of the photon 
\begin{equation}
\Lambda=2.37\times10^{-10}\ \mbox{g}.\label{eq:15.12}
\end{equation}
\pagebreak[2]

{\samepage
\begin{center}
{\bf 15. The screened fine structure constant}
\end{center}

\nopagebreak
We now consider the effects of charge screening due to
vacuum polarization on $\alpha_0$. The one-loop expression for the
fine structure constant $\alpha_\gamma$ after charge screening via
the photon is [61]
\begin{equation}
\alpha_\gamma=\frac{\alpha_0}{\displaystyle 1+\frac{\alpha_0}{3\pi}
  \,\ln\frac{\Lambda^{\prime2}}{m^2}}.\label{eq:16.1}
\end{equation}}
The cutoff $\Lambda'$ is defined according to [61]
\begin{equation}
\sum_{s=1}^S C_s\ln\lambda_s^2=-\ln\frac{\Lambda^{\prime2}}
                                   {m^2},\label{eq:16.2}
\end{equation}
where the $\lambda_s m$ are the large masses of the $S$ spinors.
The quantum mass correction Eq.\,(\ref{eq:13.4}) suggests that
the large mass of each spinor is $\Lambda$. Therefore,
\begin{equation}
\lambda_sm=\Lambda.\label{eq:16.3}
\end{equation}
Combining Eqs.\,(\ref{eq:16.2}) and (\ref{eq:16.3}), we have
\begin{equation}
\sum_{s=1}^SC_s\ln\frac{\Lambda^2}{m^2}
  =-\ln\frac{\Lambda^{\prime2}}{m^2}.\label{eq:16.2'}
\end{equation}
According to Ref.\,61,
\begin{equation}
\sum_{s=1}^SC_s=-1.\label{eq:16.4}
\end{equation}
Substituting this into Eq.\,(\ref{eq:16.2'}), we find
\begin{equation}
\Lambda'=\Lambda.\label{eq:16.5}
\end{equation}

In addition to photons producing an electron-positron pair, one might
have vacuum polarization involving the $Z^0$. To calculate the effect of
this process on $\alpha_\gamma$, we use the equation
\q
\alpha_{\gamma Z}=\frac{\alpha_\gamma}{\displaystyle 1+\frac{\alpha_\gamma}{3\pi}
\ln\frac{\Lambda_Z^{\ \,2}}{m^2}},
        \label{eq:16.6}
\qq
which is similar to Eq.\,(\ref{eq:16.1}). Here $\alpha_{\gamma Z}$
is the bare fine structure constant $\alpha_0$ after screening via the photon and $Z^0$.
The mass $\Lambda_Z$ is the classical mass of the $Z^0$ due to its weak charge. 
The classical mass $\Lambda_Z$ is determined by the classical radius equation
\q
\Lambda_Z=\frac{g^2}{2c^2r_Z},\label{eq:16.7}
\qq
where the weak charge $g$ of the $Z^0$ is given by the definition
of the weak interaction coupling constant $G_W$
\q
g^2=2^{\,5/2}\frac{G_W}{(\hbar c)^3}M_W^{\ \ 2}\hbar c,\label{eq:16.7a}
\qq
where $M_W$ is the mass of the $W$.

The procedure for determining the radius $r_Z$ of the $Z^0$ is similar to
that which determined the radius of the photon. One difference is that 
the higher-dimensional wave function for the $Z^0$ is $Y_{10}$, 
which is proportional to $\cos\theta$. The projection of this function 
onto the 4D world line or $z$-axis should therefore be proportional to 
$\cos\theta$. This, in turn, is equal to $z/r_Z$ according to the 
definition of spherical coordinates. The line segment along the 
coordinate $z$ rotates in 3DS to become the radial coordinate $r$.
Therefore the coordinate $z$ is replaced by the radial coordinate $r$.
The radial function for the $Z^0$ is
\q
R(r)=\left(\frac{3}{r_Z}\right)^{1/2}\frac{r}{r_Z}.\label{eq:16.8}
\qq
This satisfies the criteria of being proportional to $\cos\theta$ and 
normalized to one. The average value for $r$ is given by
\q
\left<r\right>=\int_0^{r_Z} R^*(r)\,rR(r)\,dr=\tfrac34\,r_Z.\label{eq:16.9}
\qq

In the velocity calculation, $\alpha_0$ appears instead of $\alpha_0^{\ 2}$
because self interactions with two vertices are allowed for the $Z^0$
as opposed to the case of the photon, where only photon-photon 
scattering (a self interaction with four vertices)
is permitted.  Here the $Z^0$'s bare fine structure constant is denoted
by $\alpha_0$. In addition,
the value of the bare fine structure constant $\alpha_0$ is given by
\q
\alpha_0=l(l+1),\label{eq:16.10}
\qq
where, from its spherical harmonic $Y_{10}$, $l=1$. 
The above considerations lead to a radius for the $Z^0$
\q
r_Z=6.7389cT.\label{eq:16.11}
\qq
Equations (\ref{eq:16.7}), (\ref{eq:16.7a}), (\ref{eq:16.11}),
and (\ref{eq:15.6}) yield 
for the classical mass of the $Z^0$ due to its weak charge
\q
\Lambda_Z=6.87\times10^{-7}\ \mbox{g}.\label{eq:16.12}
\qq

Substituting the actual mass of the electron for $m$
in Eqs.\,(\ref{eq:16.1}) and~(\ref{eq:16.6}) and 
$\alpha_0=1$ into Eq.\,(\ref{eq:16.1}) and using Eqs.\,(\ref{eq:16.5}),
(\ref{eq:15.12}), (\ref{eq:16.6}) and~(\ref{eq:16.12}),
we obtain a value for the screened fine structure constant
\q
\alpha_{\gamma Z}=\frac{1}{19.71}.\label{eq:16.13}
\qq
\pagebreak[2]

{\samepage
\begin{center}
{\bf 16. Masses of the electron, muon, tau and photon and the fine structure constant}
\end{center}

\nopagebreak
We have yet to consider the effect of the quantum charge vertex
correction on $\alpha_{\gamma Z}$. However this is
unnecessary for quantum mass corrections because it requires one
external photon line, which does not appear in the quantum mass
correction diagram. Thus, the fine structure constant used
in the quantum mass correction formula (\ref{eq:13.4})
is $\alpha_{\gamma Z}$. } To summarize, the equations that determine
the electron's mass are
\begin{eqnarray}
         \alpha_\gamma&=&\frac{\alpha_0}{\displaystyle 1+\frac{\alpha_0}{3\pi}
         \ln\frac{\Lambda^2}{m^2}},\label{eq:17.1a}\\ 
         \alpha_{\gamma Z}&=&\frac{\alpha_\gamma}{\displaystyle 1+\frac{\alpha_\gamma}{3\pi}
         \ln\frac{\Lambda_Z^{\ \,2}}{m^2}},\label{eq:17.1b}\\
         m&=&\Lambda\,\exp\!\left(\frac14-\frac{2\pi}{3\alpha_{\gamma Z}}
         \right),\label{eq:17.1c}
\end{eqnarray}
where $\alpha_0=1$ and $\Lambda$ and $\Lambda_Z$ are given
by Eqs.\,(\ref{eq:15.12}) and~(\ref{eq:16.12}), respectively.
Solving the three Eqs.\,(\ref{eq:17.1a})--(\ref{eq:17.1c})
for the three unknowns
$\alpha_\gamma$, $\alpha_{\gamma Z}$ and $m$, we obtain
\q
m=\Lambda^5\Lambda_Z^{\ \,-4}\,\exp\!\left(\frac94-\frac{6\pi}
{\alpha_0}\right).\label{eq:17.2}
\qq

The value for $m$ obtained from Eq.\,(\ref{eq:17.2}) is
$1.16\times10^{-4}$
MeV, which is much less than the actual electron
mass of 0.511 MeV. This is because we have calculated the
mass of the electron due to its electric charge. To this 
we must add the contribution from its weak charges.

The electron has two weak charges:  that from weak isospin
and one from an explicit weak interaction term in the 6D curvature
scalar $\widehat R$ of 3D time.  
The mass of the electron due to its weak charges is obtained from
Eq.\,(\ref{eq:17.2}), but with $\alpha_0$
determined by the electron's weak charge from its
weak isospin $I$ or its weak charge from the 6D curvature
scalar $\widehat R$. The electron's weak charge $g_{\widehat R}$ 
from $\widehat R$ in Part~I is given by Eq.\,(\ref{eq:7.3}) and
is $-4.75766\sqrt{\hbar c}$. Therefore
\q
\alpha_0=\frac{g_{\widehat R}^{\ \ 2}}{4\pi\hbar c}
          =1.8013.\label{eq:17.3}
\qq
Substituting this into Eq.\,(\ref{eq:17.2}), we obtain for the weak mass
of the electron 0.510 MeV, which has an error of three tenths
of one percent.
  
The electron's remaining weak charge $g_I$ due to its weak isospin
$I=L=\left[l(l+1)\right]^{1/2}\hbar$ is 
$\left[l(l+1)4\pi\hbar c\right]^{1/2}$, where from Table~1,
$l=\frac12$ for the electron. Exchanging $g_I$ for $g_{\widehat R}$
in Eq.\,(\ref{eq:17.3}) we find the mass of the electron due to
its weak isospin is negligible.
Thus, the prediction for the electron's mass stands at an accurate
0.510 MeV. We have determined
our first previously unexplained parameter --- the mass of the electron.

The masses of the muon and tau are calculated in a manner similar
to that of the electron. Their electromagnetic masses remain the 
same because the electromagnetic $\alpha_0$ stays at unity. Their weak masses, however,
increase drastically because their weak $\alpha_0$'s  due to 
isospin increase from $l(l+1)=\frac12\left(\frac12+1\right)$
for the electron to $\frac32\left(\frac32+1\right)$ for the muon
and $\frac52\left(\frac52+1\right)$ for the tau. 
The mass depends on the weak $\alpha_0$ exponentially. 
It is this exponential dependence of the mass upon the 
weak charge that causes the muon to be so much more massive
than the electron.

Like the electron, the mass of the muon will be given by the sum
of its masses due to its weak charges from isospin and from $\widehat R$.
The weak charge from isospin yields $\alpha_0 = l(l+1)$, where $l=3/2$.
We have $\alpha_0=3.75$.
Equation (\ref{eq:17.2}) yields
$m=117.3$ MeV, which is 11\% higher than the actual
muon mass of 105.7 MeV.
The weak charge for the muon $g_{\widehat R}$
from the 6D curvature
scalar is the same as that for the
electron [Part~I]. The mass contribution from this weak charge for the muon
is the same as that for the electron
(one electron mass) and is therefore negligible.
Thus its mass is primarily due to isospin and is predicted to be 117.3 MeV.
The mass of the muon is the second previously unexplained parameter
of the SM to be explained by 3D time.

(The mass of the muon comes from its motion on the HD sphere. 
At first glance this appears to be something apart from the curvature scalar $\w R$,
but it is $\w R$ that puts a moving muon on the HD sphere.
This follows from its HD angular momentum representation $\left |\frac32,-\frac12\right >$.)

Like the muon, the mass of the tau due to its weak charge from
$\widehat R$ is one electron mass and is negligible.
Therefore, its weak $\alpha_0$ is from isospin and is given by $l(l+1)$,
where $l=\frac52$. Its calculated mass is 2072 MeV, which is 16\%
higher than its actual mass of 1784 MeV.
This is the third previously unexplained parameter to be explained by 3D time.

The error in the calculation of masses for the muon and tau is substantially 
larger than that for the electron. This is where the one loop approximation in the 
quantum mass and charge correction equations begins to break down.
One would expect this of a first-order, one-loop 
approximation to a series in powers of fractions of $\alpha_0$. The higher-order 
terms do not drop off as fast for increasing $\alpha_0$, resulting in increased 
error for the one-loop approximation.

The value of the fine structure constant $\alpha$ 
is determined from
$\alpha_{\gamma Z}$ by the final quantum charge correction,
known as the vertex correction.
The one-loop expression for the fine structure constant $\alpha_V$ after
the vertex quantum correction is [61]
\q
\alpha_V=\frac{\alpha_{\gamma Z}}{\displaystyle \left(1+\frac{\alpha_{\gamma Z}}{3\pi}
  \ln\frac{m}{\mu}\right)^2},\label{eq:17.4}
\qq
where $\alpha_{\gamma Z}$ is given by Eq.\,(\ref{eq:16.13}),
$m$ is the mass of the electron,
and $\mu$ is the actual mass of the photon. 

To calculate $\mu$, we adapt the mass formula Eq.\,(\ref{eq:17.1c}) for use by the photon. 
We have 
\q
\mu=\Lambda\,\exp\!\left(\frac14-\frac{2\pi}{3\beta_{\gamma Z}}\right),\label{eq:17.4c}
\qq
where
\q
\beta_{\gamma Z}=\alpha_{\gamma Z}^{\ \ 2}=\left(\frac{1}{19.71}\right)^2.\label{eq:17.5}
\qq

Substituting Eq.\,(\ref{eq:17.5}) into Eq.\,(\ref{eq:17.4c}), we obtain
\q
\mu=1.3317\times10^{-363}\ \mbox{g}.\label{eq:17.6}
\qq
Because $\mu\ne0$ the quantum vertex
correction in Eq.\,(\ref{eq:17.4}) is not infinite. Equation (\ref{eq:17.4})
then yields
\q
\alpha_V=\frac{1}{525.44}.\label{eq:17.7}
\qq
This value for the fine structure constant is substantially less than the observed value of 1/137.
This is because we have calculated the value for the fine structure constant due to
vacuum polarization when the photon or $Z^0$ emits an electron-positron pair.
To this we must add the contributions to $\alpha$ when a muon-antimuon pair and a tau-antitau
pair are emited. We must recalculate $\alpha_V$ twice more and add all three results.
The recalculations are done by substituting $\alpha_0 = 1$ into Eq.\,(\ref{eq:17.1a})
as we did for the electromagnetic contribution to the mass of the electron. 
This is done because if one wants to wind up with the 
quantum-corrected fine structure constant for electromagnetism one must start off with the 
bare fine structure constant for electromagnetism. In addition, the mass of the muon is used in
Eqs.\,(\ref{eq:17.1a}), (\ref{eq:17.1b}) and (\ref{eq:17.4}) instead of the mass of the electron. 
It is only through the muon's mass that the muon affects the calculation for $\alpha_V$.
Then $\alpha_\gamma$ is substituted into
Eq.\,(\ref{eq:17.1b}) and $\alpha _{\gamma Z}$ is determined. Next, the photon mass
is calculated from Eq.\,(\ref{eq:17.4c}) and substituted into Eq.\,(\ref{eq:17.4}).
This procedure is repeated for the tau.
The values for $\alpha_V$ from the electron, muon and tau are added. We have
\begin{eqnarray}
         \alpha&=&\alpha_{V e}+\alpha_{V \mu}+\alpha_{V \tau}\label{eq:17.8}\\ 
         &=&\frac{1}{525.44}+\frac{1}{379.62}+\frac{1}{314.67}\label{eq:17.9}\\
          &=&\frac{1}{129.61}.\label{eq:17.10}
\end{eqnarray}
Note that adding $\alpha_{V \mu}+\alpha_{V \tau}$ to $\alpha_{V e}$ is about the
same as multiplying $\alpha_{V e}$ by $\sqrt{4\pi}$. We have previously used the factor of
$\sqrt{4\pi}$ instead of the addition of $\alpha_{V \mu}+\alpha_{V \tau}$ because 
this is more accurate.

The value 1/129.61 for the fine structure constant for the electron for electromagnetism is roughly 
six percent too large. The error is due to the one loop approximations. More loops means
more vacuum polarization and a smaller $\alpha$. Mathematically, this means more terms
in the denominators of Eqs.\,(\ref{eq:17.1a}), (\ref{eq:17.1b}) and (\ref{eq:17.4}).
These denominators are the first two terms of infinite series in $\alpha_0$, $\alpha_\gamma$
and $\alpha _{\gamma Z}$, respectively. The larger denominators mean smaller values for
$\alpha_\gamma$, $\alpha_{\gamma Z}$ and $\alpha_V$, respectively. This leads to a smaller
value for $\alpha$.

During the calculation for $\alpha_{V}$, different values for the photon mass $\mu$ 
were obtained for each of the electron, muon and tau cases. Apparently, each of these is a
contribution to the total photon mass. The contributions from the electron and muon cases
are negligible compared to that of the tau. The resulting photon mass is $\mu=1.55\times10^{-250}\ \mbox{g}$.
This mass for the photon is effectively zero.
It is far below the experimental limit of
$4\times10^{-48}\ \mbox{g}$. The mass of the photon is generally
thought to be zero, although no one knows why.
There is no currently accepted explanation for this mass.
The classical theory of 3D time showed why it is zero.
Now we have a complete and consistent explanation for the mass of the photon,
right down to the fact that it must be small but nonzero 
in order to eliminate infrared divergences from QED.
This mass is the fifth previously unexplained parameter
of the SM to be explained by 3D time.

\pagebreak[2]

{\samepage
\begin{center}
{\bf 17. Masses of the neutrinos}
\end{center}

\nopagebreak

We will now derive the masses of the neutrinos.
Solving Eq.\,(\ref{eq:15.7}) for
the integer and substituting $c=\frac12$ and $m=\frac12$ for the
neutrino, we find the charge on the neutrino is zero. This means
neutrinos do not couple to the photon. Therefore, their electromagnetic
masses are zero.} Neutrinos have the same weak charges 
$g_I$ and $g_{\widehat R}$ as their charged leptonic counterparts.
However, since the photon is absent, their weak $\alpha_0$ 
is screened only via the $Z^0$. Thus, there will be only one
vacuum polarization equation.  In addition, the electromagnetic
cutoff $\Lambda$ will be absent from the quantum mass correction
equation. It is replaced by a weak cutoff described below.

The equations that determine the masses of neutrinos are
\begin{eqnarray}
  \alpha_Z&=&\frac{\alpha_0}{\displaystyle 1+\frac{\alpha_0}{3\pi}
             \ln\frac{\Lambda_Z^{\ \,2}}{m_\nu^{\ \,2}}},
                      \label{eq:18.1a}\\
         1&=&\frac{3\alpha_Z}{4\pi}\left(\ln
             \frac{\Lambda_Z\Lambda_Z'}{m_\nu^{\ 2}}+\frac12\right),
                      \label{eq:18.1c}
\end{eqnarray}
where Eq.\,(\ref{eq:18.1c}) is Eqs.\,(\ref{eq:13.1}) and~(\ref{eq:13.2})
after $m$ cancels when $m_0=0$.
One of the $\Lambda$'s in this equation must be
$\Lambda_Z'=\Lambda_Zm_\nu^{\ 2}/M_W^{\ \ 2}$. In other words,
$\Lambda_Z'$ is simply $\Lambda_Z$ with the $W$ mass squared replaced
by the neutrino mass squared. This is necessary in order to allow for
the weakness of the weak interaction where the neutrino emits a $Z^0$ at the
first vertex of the quantum mass correction diagram.
This process is a weak one due to the large mass of the $Z^0$.
However, all subsequent processes described by
the vacuum polarization and quantum mass correction diagrams should
not be `tagged' as weak because the $Z^0$ has already been produced ---
the process is not further hindered by the production of a more massive particle.
This means that all $\Lambda$'s in the 
vacuum polarization and mass correction equations besides
the $\Lambda_Z'$ mentioned above must be $\Lambda_Z$.

In the electron's quantum mass correction
equation, the classical mass of the electron was given by the sum of its
classical electromagnetic and weak masses, but the classical weak mass
$\Lambda_Z'=\Lambda_Zm_e^{\ 2}/M_W^{\ \ 2}=2.78\times10^{-17}$ g
was negligible compared to the electromagnetic
$\Lambda= 2.37\times10^{-10}$ g.
Now that the photon's effects are absent from the quantum mass
equation, $\Lambda$ is replaced by $\Lambda_Z'$.

Note that the neutrino mass $m_\nu$ cancels in
Eq.\,(\ref{eq:18.1c}). The relative smallness of $\Lambda_Z'$ is the
reason why neutrinos have such small masses. And the only reason we need
to consider $\Lambda_Z'$ is because the neutrino has zero charge.
Therefore, neutrinos have such small masses because they have no charge,
as one might expect.

The solution of Eqs.\,(\ref{eq:18.1a}) and~(\ref{eq:18.1c}) is
\q
   m_\nu=\Lambda_Z^{-1/8}\left(\frac{\Lambda_Z'}{m_\nu^{\ 2}}\right)^{-9/8}
   \exp\!\left(\frac{3\pi}{2\alpha_0}-\frac{9}{16}\right).\label{eq:18.2}
\qq
Because the fraction involving $\alpha_0$ in the exponential in
Eq.\,(\ref{eq:18.2}) does not have a minus sign in front of it as
it does for charged leptons, the masses of neutrinos decrease with
increasing $\alpha_0$. Thus, the smaller of $\left|g_I\right|$ or $\left|g_{\widehat R}\right|$
instead of the larger will yield the masses of the neutrinos.
The larger of these two weak charges will yield a mass that is negligible.

Each neutrino will have the same weak charges $g_I$ and $g_{\widehat R}$
as its charged leptonic counterpart.  For the electron neutrino
$g_I=3.07\sqrt{\hbar c}$ (leading to $\alpha_0=3/4$) and
$g_{\widehat R}=-4.758\sqrt{\hbar c}$ (leading to $\alpha_0=1.8013$).
Substituting these separately into Eq.\,(\ref{eq:18.2}),
we find that the mass of the electron neutrino is due to $g_I$ and is
$6.16\times10^{-7}$ eV, which is
well below the experimental limit of 20 eV. The masses of the muon 
and tau neutrinos are determined from $g_{\widehat R}=-4.758\sqrt{\hbar c}$.
They are the same and are $1.57\times10^{-8}$ eV,
far lower than their experimental limits of 0.2 MeV and 35 MeV,
respectively.
The masses of the neutrinos are the sixth, seventh and eighth previously 
unexplained parameters of the SM determined by 3D time.
\pagebreak[2]

{\samepage
\begin{center}
{\bf 18. New fermions}
\end{center}

\nopagebreak
According to Table~1, other leptons exist in the muon's and tau's multiplets
besides the known particles. These particles turn out to have properties
identical to the known particles but have different quantum numbers.}
Apparently, we have been detecting them but have been mistaking them
for the muon, tau, and their neutrinos. The usual muon is symbolized by $\mu^-$.
The new charged particle in the muon's multiplet has the symbol $\mu_2^+$. 
Table~1 gives the quantum numbers for the new particle as $l=\frac32$ and $m=\frac32$.
According to the Gell-Mann relation (\ref{eq:15.7}), its bare
charge quantum number is $Q=1$.  The $\mu_2^+$
has the same electromagnetic and weak
bare fine structure constants as the usual
muon $\mu^-$. Therefore, it has the same mass as the muon. 
Because its charge is reversed it is similar to
the antimuon ${\overline\mu}^{\,+}$.
However, it is not the same as the antimuon because
its quantum number $m=\frac32$ is different than
the antimuon's $m=\frac12$.  Because this particle has the same
mass as the usual muon,
there is no reason why it should not have been
detected already. I believe we have been detecting it all along
but have been mistaking it for the antimuon.

The particle with $l=\tfrac32$ and $m=-\tfrac32$ in Table~1 has
$Q=-2$ according to the Gell-Mann relation (\ref{eq:15.7}).
Therefore, $\alpha_0=Q^2=4$. This would put its electromagnetic mass
at about the muon's weak mass because the muon's weak $\alpha_0=3.75$.
Remember that a spinor's mass comes from its charge 
and that a particle may have more than one type of charge.
The mass contributions from the different types of charge add.
Because this particle has $l=\frac32$, which is the same value for $l$ as the muon, its
weak mass would be that of the muon. Its total mass would therefore, be
about double that of the muon (266 MeV to be exact).
However, a lepton with this mass is not observed.
This must be due to the fact that the 
electromagnetic $\alpha_0$ is too large for the higher-loop
series [61] in terms of
$\alpha_0$ to converge. This electromagnetic series for
vacuum polarization becomes infinite,
resulting in an infinite denominator for $\alpha_\gamma$. The parameter
$\alpha_\gamma$ is therefore zero. This means the charge is 
completely screened by the photon. In effect, the particle has no charge
and does not couple to the photon. Therefore, it is a neutrino.
Because it is in the muon's multiplet, it is a muon neutrino. It would have
the same mass as, but be different than, the ordinary muon neutrino,
which has $m=\frac12$.

The new leptons in the tau's multiplet have $l=\frac52$ and
$m=\pm\frac52,\pm\frac32$, according to Table~1.
These $m$ quantum numbers distinguish the new particles from
the tau and its neutrino, which have $m=\pm\frac12$.
The bare charge quantum numbers of these new particles are
$Q=-3,-2,1,2$. By the same reasoning used above for the muon's
multiplet, the particle with $Q=1$ has properties similar to
the antitau, while the others are new tau neutrinos.

Fermions in the next spinor multiplet beyond the tau's have $l=\frac72$
and $\frac72 \geq m \geq -\frac72$. Because none of these particles
are observed, the weak $\alpha_0$ must be too large
for the higher-loop series
in terms of this $\alpha_0$ to converge.
Because this series in terms of the weak $\alpha_0$ does not
converge, the weak charges of these $l=\frac72$ particles are zero,
by the same argument used above for the vanishing of the electric charge.
Therefore, they do not couple to the $Z^0$.  This means they would not
affect the decay width, which is the test for their existence,
of the $Z^0$. This is the reason why there appears to be only
three families of leptons.

The particles with $Q=0$ or $|Q|\geq 2$ also have no electric charge
and therefore, have no electromagnetic masses because
$\alpha_\gamma=0$. These particles would acquire a
small mass through their superweak couplings to $A_\mu^{lm}$,
$l\geq 2$. These superweak masses should be smaller than
ordinary neutrino masses, which are small due to the weakness of
the weak interaction.

The particle with $Q=1$ has electric charge but no weak charge and
is screened by the photon only. The equations that determine the mass
of this particle are
\begin{eqnarray}
  \alpha_\gamma&=&\frac{\alpha_0}{\displaystyle 1+\frac{\alpha_0}{3\pi}
  \ln\frac{\Lambda^2}{m^2}},\label{eq:19.1a}\\
         m&=&\Lambda\exp\!\left(\frac14-\frac{2\pi}{3\alpha_\gamma}\right).
         \label{eq:19.1b}
\end{eqnarray}
The solution of Eqs.~(\ref{eq:19.1a}) and~(\ref{eq:19.1b}) is
\q
m=\Lambda\exp\!\left(\frac{9}{20}-\frac{6\pi}{5\alpha_0}
    \right).\label{eq:19.2}
\qq
Substituting $\alpha_0=1$ into Eq.~(\ref{eq:19.2}),
we find $m=4.81\times10^{9}$ TeV.

Because their electric and weak charges are the same, spinors with
$l>\frac72$ will have masses similar to those with $l=\frac72$.
That is, particles with $Q=0$ or $|Q|\geq 2$ will have neutrino-like masses
and those with $|Q|=1$ will have the mass $4.81\times10^{9}$ TeV.

\pagebreak[2]

{\samepage
\begin{center}
{\bf 19. Elementary particle interactions}
\end{center}

\nopagebreak
Just as a given particle decays only into certain decay products,
a given ket may only have certain other kets as its factors.
In fact, the relationship is one-to-one, with the ket for each
particle determining which decays can take place.
For example, one may ask why the $W^+$ decays into the particles
it does, instead of the decay products for the $W^-$
or the $Z^0$.  To be sure, these decays can be ruled out by
the law of conservation of charge, but then one must ask why
the law of conservation of charge holds.  I claim this law follows
from the rules for combining kets given by
Clebsch-Gordan coefficients. These rules uphold the law of
conservation of angular momentum.}

One may know which
elementary particle interactions can take place by
determining whether or not the relevant Clebsch-Gordan coefficient is
zero. An elementary particle interaction can take place if the
coefficient is not zero. This is the case if
and only if $m=m_1+m_2$ and $|l_1+l_2| \geq l \geq |l_1-l_2|$. 
I claim the relation $m=m_1+m_2$ is
the law of conservation of charge.
In terms of the bare charge quantum number $Q$ for vectors,
we have $Q=m$; in addition, for spinors
$Q=m-\tfrac12$, and for antispinors $Q=m+\tfrac12$.

A short cut to obtaining the spinor expansion of a vector is to
write the vector's ket in terms of half-odd integral $l$ kets
using Clebsch-Gordan coefficients. One then writes the particle
symbols of the spinors with the quantum numbers of the spinor
kets below the spinor kets. The quantum
numbers of spinors are given in Table~1. Remember that the
first spinor ket in each pair is associated with an antiparticle
by convention.

All of this is made clear by a simple example:  The ket 
$|11\rangle$ is equal to the product of kets
$\left|\tfrac12\,\tfrac12\right>\left|\tfrac12\,\tfrac12\right>$
because the relevant Clebsch-Gordan coefficient is not zero.
It is one. We have
\begin{equation}
|11\rangle=
\left|\tfrac12\,\tfrac12\right>\left|\tfrac12\,\tfrac12\right>.
\label{eq:21.2}
\end{equation}

As in Sec.\,8, we identify a spinor with a bar
over it with the first half-odd integral $l$ ket.
After identifying $|11\rangle$ above with the $W^+$, the first 
$\left|\tfrac12\,\tfrac12\right>$ with ${\overline e}^{\,+}$
(a positron) and the second
$\left|\tfrac12\,\tfrac12\right>$ with $\nu_e^{\,0}$
(a neutrino), we have the decay
$W^+\to {\overline e}^{\,+}+\nu_e^{\,0}$. Another way of looking at this is
that the above interaction is justified
by substituting a spinor expansion for one factor of $W_\mu^{+}$
in its mass term in the Lagrangian, thereby generating an interaction term
as described in Part~I.

The $W^+$ (ket $|11\rangle$)
cannot decay into two neutrinos (kets $\kk\k$) because
the Clebsch-Gordan coefficient for this process is zero.
Similarly, the $W^+$ cannot decay into two electrons, two positrons,
a positron and an electron, or an electron and a neutrino because
the Clebsch-Gordan coefficients for all these processes are zero.
These violate $m=m_1+m_2$, the law of conservation of charge.

There is one more rule for obtaining
nonzero Clebsch-Gordan coefficients involving leptons.
As it stands, there is nothing to prevent a ket $|lm\rangle$ from being
expanded in terms of kets with different values for half-odd
integral $l_1$ and $l_2$.
For example, if $l=1$, $l_1=\tfrac12$ and $l_2=\tfrac32$, then it would
appear one could have
$|11\rangle=
\sqrt{\tfrac{1}{4}}
\left|\tfrac12\,\tfrac12\right>\left|\tfrac32\,\tfrac12\right>
-\sqrt{\tfrac{3}{4}}
\left|\tfrac12\,,-\tfrac12\right>\left|\tfrac32\,\tfrac32\right>$,
with the corresponding interactions $W^+\to {\overline e}^{\,+} 
+ \nu_{\mu}^{\,0}$ for the first pair of kets and
$W^+\to  {\overline\nu}_{e}^{\,0} +\mu^+_2$ for the second. 
Both of these decays violate electron and muon number conservation. 
One needs a reason why this cannot occur --- it contradicts observation.
The reason is as follows. All spinors in 3D time are ultimately
derived from the photon because this is the only vector that is null.
As explained in Ref.\,(45),
this is a requirement for writing a vector in terms of two spinors.
Now one cannot expand the ket for the photon $|00\rangle$ in terms
of kets with different values for $l_1$ and $l_2$ because then the
minimum value for $l$ will be greater than zero, while the photon
must have $l$ equal to zero. The minimum value for $l$ in the above
example was $|\tfrac12\,-\tfrac32|=1$, which is not zero.
(Remember the minimum value for $l$ is given by $|l_1-l_2|$ from the
relation $|l_1+l_2| \geq l \geq |l_1-l_2|$). This means that we must have
$l_1=l_2$ in order to achieve $l=0$ and that $|00\rangle$ can only be 
expanded in terms of kets from one family of leptons at a time. 
This means one could expand
$|00\rangle$ in terms of $l_1=l_2=\tfrac12$ kets or $l_1=l_2=\tfrac32$
kets but not both in the same expansion. In effect, only one family of
leptons may exist at one time.  This adds the requirement
$l_1=l_2$ to the above rules for obtaining nonzero Clebsch-Gordan
coefficients involving leptons and ensures that electron and muon
number is conserved.

The elementary particles are described by the Lagrangian for 3D time.
The Lagrangian for 3D time, in turn, determines 6D spacetime.
It may be that conjugation of charge, parity and time,
known as CPT conjugation, is an invariance
because reversal of all six dimensions leads to an identical spacetime.
\pagebreak[2]

{\samepage
\begin{center}
{\bf 20. Masses of the W and Z}
\end{center}

\nopagebreak
One source of mass in 3D time is the appearance of explicit
mass terms in the 6D curvature scalar $\w R$.
For spinors, the graviton and the photon this mass is zero, while for
the $Z^0$ and $W^\pm$ these masses [from Part~I] are 1.9109\,$\hbar/c^2T$
and 1.6293\,$\hbar/c^2T$, respectively. (Note that $cT$ is the Planck
length and $\hbar/c^2T$ is the Planck mass.)} To these masses we should
add the mass due to the particle's charges. One obtains this type of
mass by substituting the particle's charge into the classical radius equation.
The predicted masses of the $W$ and $Z$ without these masses will
turn out to be close enough to the
actual values to allow us not to do this calculation.
Therefore, the masses of the $Z^0$ and $W^\pm$ are those from
$\w R$ given in Part~I.

Now that we have the Planck masses for the $W^\pm$ and $Z^0$,
the question arises, ``How do we reduce these masses to their
observed values?" We noted above that a vector can be
considered to be a combination of two spinors. Thus, we can
try the quantum mass correction developed for spinors for vectors.
Both the wave functions for the $Z^0$ and the $W^\pm$ are made
up of those for the electron and electron neutrino.  From its
spinor expansion, one can say that the $Z^0$ is one half an
electron-positron pair and one half a neutrino-antineutrino pair.
The factors of one half arise from the Clebsch-Gordan coefficients
of $1/\sqrt2$ multiplying each of the two pairs of spinor kets.
The mass of the neutrino-antineutrino pair is negligible compared
to that of the electron-positron pair. Thus, the $Z^0$'s mass
contributions are from the electron-positron pair and are
one-half of twice the electron's mass or simply one electron mass.
The $W^+$ is a positron-neutrino pair and the $W^-$ is an
electron-antineutrino pair. Because the neutrino masses are
negligible, each $W$ mass is based on one electron mass.

The quantum mass corrections  of the $Z^0$ and $W^\pm$
will be that of the electron because the masses are based on one electron
mass. Therefore,
\q
M_W=\exp\!\left(\frac14-\frac{2\pi}{3\alpha_{\gamma Z}}\right)
  \Lambda(W_\mu^\pm),\label{eq:20.3} 
\qq
where we have denoted the above-mentioned Planck mass of the
$W$ by $\Lambda(W_\mu^\pm)$. In addition,
$\alpha_{\gamma Z}=1/19.71$, which is the value that yields the
mass of the electron in Eq.\,(\ref{eq:17.1c}).
Dividing Eq.\,(\ref{eq:20.3}) by Eq.\,(\ref{eq:17.1c})
when  $\alpha_{\gamma Z}$ equals $1/19.71$, we obtain
\q
\frac{M_W}{m_e}=\frac{\Lambda(W_\mu^\pm)}{\Lambda}.\label{eq:20.4}
\qq
Using Eq.\,(\ref{eq:20.4}) to solve for the mass of the $W^\pm$, we find
$M_W=76.48$ GeV. Similarly,
\q
\frac{M_Z}{m_e}=\frac{\Lambda(Z_\mu^0)}{\Lambda},\label{eq:20.5}
\qq
where $\Lambda(Z_\mu^0)$ is the classical Planck mass
of the $Z$. This leads to a $Z^0$ mass of 89.70 GeV. 

These masses are in fairly good 
agreement with experiment [64], where $M_W=80.40\pm0.2$ GeV and
$M_Z=91.1867\pm0.0021$ GeV. The errors are presumably due to
lack of consideration of the mass contributions from their charges.
The SM does not explain the masses of the W and Z.
It derives them from the weak interaction
coupling constant $G_W$, which comes from experiment.
3D time explains the masses of these two particles from scratch,
allowing, instead, one to derive $G_W$
from them.  These are parameters nine and ten of the SM.
These are the final parameters of the SM to be derived for now.

It should not be surprising that we have arrived close to correct values for the 
masses and charges of the elementary particles. If your classical theory is correct,
then applying quantum mechanical corrections to it should result in the correct
values for physical quantities.

\pagebreak[2]

{\samepage
\begin{center}
{\bf 21. Predicted intermediate vector bosons}
\end{center}

\nopagebreak
The vector $A_\mu$ in 3D time
is expanded in terms of spherical harmonics $Y_{lm}$
\q
A_\mu=\sum_{l=0}^\infty \sum_{m=-l}^l A_\mu^{lm}Y_{lm}.\label{eq:21.1}
\qq
We neglect the strong interactions in what follows.
The $l=0$ coefficient $A_\mu^{00}$ in this expansion is the photon.
The $l=1$ coefficients $A_\mu^{1m}$ are the $W$ and $Z$.
The $l=2$ and $l=3$ coefficients $A_\mu^{2m}$ and $A_\mu^{3m}$
are vectors for two new superweak interactions.}

The procedures used to obtain the masses of the
$A_\mu^{2m}$ from the spinors of which they are composed are
the same as those described for the $A_\mu^{1m}$ in the last section.
Remember that in order to obtain nonzero Clebsch-Gordan coefficients
for leptons, $l_1=l_2$, which means that one cannot expand
$l=2$ kets in terms of $l_1=\frac32$ and $l_2=\frac12$ kets
as one might expect, but rather kets with $l_1=l_2=\frac32$ only.
The wave function of the vector $A_\mu^{2m}$ is composed of those of
the muon and the muon neutrino.  The mass contributions of these particles
add. The neutrino contribution is negligible.  An exact prediction
for the mass of $A_\mu^{2m}$ can be made if we multiply the classical
mass ratio $\Lambda(A_\mu^{2m})/\Lambda$ by the mass of the muon
[similar to Eq.\,(\ref{eq:20.4})].

From $\Lambda(A_\mu^{20})=3.4032\,\hbar/c^2T$ (where $\hbar/c^2T$
is the Planck mass), we obtain for $A_\mu^{20}$ a mass of 33.04 TeV.
Similarly, $\Lambda(A_\mu^{2,\pm1})=3.2159\,\hbar/c^2T$, which leads to a
mass of 31.22 TeV for $A_\mu^{2,\pm1}$. The classical mass
$\Lambda(A_\mu^{2,\pm2})=0.46994\,\hbar/c^2T$ yields the actual mass
4.56 TeV for the $l=2$, $m=\pm2$ particle. These are approximate masses.
The actual masses of these particles could be as much as 5\% higher or more
due to the mass contributions from their weak charges. 

The vectors $A_\mu^{3m}$ are composed of a muon and a muon neutrino.
Their masses are obtained by multiplying
their classical mass ratios by the mass of the muon. We have
$\Lambda(A_\mu^{30})=3.9957\,\hbar/c^2T$,
$\Lambda(A_\mu^{3,\pm1})=3.0315\,\hbar/c^2T$,
$\Lambda(A_\mu^{3,\pm2})=2.8175\,\hbar/c^2T$ and
$\Lambda(A_\mu^{3,\pm3})=0.75385\,\hbar/c^2T$.
The masses of these particles are 38.79 TeV, 29.43 TeV, 27.36 TeV and
7.32 TeV, respectively. Again, these are approximate masses.

The electric charges of the $A_\mu^{2m}$ are not, in general, given by
their quantum number $m$.  This is because $m$ is the vector's
bare charge quantum number, which is not necessarily its
actual charge quantum number.  The vector's actual charge
quantum number is given by
the sum of the actual charges of the fermions in its spinor expansion.
This is because the vector may be thought to come about from the
combination of the spinors from which it is composed.
Another way of looking at this is that a vector and its spinor expansion
must be mathematically and physically identical.  For example, consider
the $Z^0$.  From its spinor expansion, we find the $Z^0$ is composed of
one-half an electron-positron pair and one-half a neutrino-antineutrino
pair.  Thus, the charge of the $Z^0$ is given by 
$\frac12(-1+1)+\frac12(0+0)$, which equals zero.
Had one of the numbers in one set of the parentheses been different, the $Z^0$
would have wound up with a different charge ---
and a fractional one at that. One of the numbers will
be different for the case of the spinor with bare charge quantum number $-2$
and actual charge quantum number 0.  This spinor is in the
muon's multiplet and is described in Sec.\,18 (it has $Q=-2$). Therefore,
instead of having a $-2$ here we will have a 0.  This is why the electric charge of
$A_\mu^{lm}$ is not, in general, given by $m$ and why it may have a
fractional charge.

The charges of the next two generations of intermediate vector bosons ---
the $A_\mu^{lm}$ where $l=2$ or 3 are as follows.  
The charge of the superweak vector $A_\mu^{2,\pm2}$
is $\pm1$.  The vector $A_\mu^{20}$
is neutral, while $A_\mu^{2,\pm1}$ has a charge of $\pm\frac34$.
The charge of the superweak vector $A_\mu^{3,\pm3}$
is $\pm1$.  The vector $A_\mu^{30}$
is neutral, while $A_\mu^{3,\pm1}$ has a charge of $\pm\frac35$.
The charge of $A_\mu^{3,\pm2}$ is $\pm1$.
Remember, one can only detect a superweak vector's decay products
so these fractional charges can not be observed. (Each of its decay 
products has a charge of $-1$, 0, or $+1$.)

\pagebreak[2]

{\samepage
\begin{center}
{\bf 22. Predictions and the so-called Higgs particle}
\end{center}

\nopagebreak
I claim that the scalar particle announced as discovered at the LHC on Wednesday, July 4, 2012
is not the Higgs boson but merely an ordinary scalar particle with no special properties.
It is $\w g^{\,55}=1+A^\m A_\m$. Because it has spin 0
it is a boson and is therefore the carrier of a fifth force. The force is weak
because of the large mass of the carrier.} We rename it $\w g^{\,55}=\Phi$.
The field $\Phi$ has a nonzero vacuum expectation value because if $A_\m=0$ then
$\Phi=1$. However, according to 3D time, there is no Higgs mechanism,
so this point is moot.

To calculate the terms in the Lagrangian that apply to $\Phi$, we define
the 6D metric tensor in terms of 4D quantities as follows
\begin{equation}
\left(
\begin{array}{ccc}
\w g_{\mu\nu}&\w g_{\mu5}&\w g_{\mu6}\\
\w g_{5\nu}&\w g_{55}&\w g_{56}\\
\w g_{6\nu}&\w g_{65}&\w g_{66}
\end{array}
\right)=\left(
\begin{array}{ccc}
\delta_{\mu\nu}+A_\mu A_\nu&A_\mu&0\\
A_\nu&1&0\\
0&0&1
\end{array}
\right).\label{eq:22.2}
\end{equation}

The contravariant metric tensor is
\begin{equation}
\left(
\begin{array}{ccc}
\w g^{\,\mu\nu}&\w g^{\,\mu5}&\w g^{\,\mu6}\\
\w g^{\,5\nu}&\w g^{\,55}&\w g^{\,56}\\
\w g^{\,6\nu}&\w g^{\,65}&\w g^{\,66}
\end{array}
\right)=\left(
\begin{array}{ccc}
\delta^{\mu\nu}&-A^\mu&0\\
-A^\nu&
\Phi&0\\
0&0&1
\end{array}
\right).\label{eq:22.3}
\end{equation}

These values for the 6D metric tensor are substituted into the Christoffel symbols,
which are substituted into the curvature scalar. This calculation should be much quicker 
than that for the free-$A_\m$ field but slightly longer than that for the free-$g_{\m\n}$ field.
The mass of $\Phi$
will depend on the coefficient of its mass term derived from this calculation.

The vectors $A_\mu^{lm}$ can be expanded in terms of
spin 1/2 combinations of three quarks. This explains where baryons come from.
The scalar $\Phi$ may be expanded in terms of spin $0$ combinations
of quarks and antiquarks. This is where mesons originate.

The predictions 3D time makes are as follows.
3D time predicts no supersymmetric particles. Instead, it predicts new 
intermediate vector bosons and elementary tensor particles. One of the tensor particles
has a mass of approximately 58 GeV. A search~[65] for scalars with similar decay products
has already been conducted throughout the mass range 65--600 GeV. Nothing was found 
besides the so-called Higgs boson. If the search is
widened, the tensor particle, named $g_{\m\n}^{10}$, may be detected.
The least massive of the vectors is $A_\mu^{2,\pm2}$ with an approximate mass of
4.56 TeV. A search [66] for such particles at center-of-mass energy 8 TeV
revealed nothing up to 3.24 TeV. The LHC will begin operating at 
center-of-mass energy 13 TeV in 2015 allowing detection of such particles up to
the mass 5.26 TeV. Perhaps it could be found then.
\pagebreak[2]

{\samepage
\begin{center}
{\bf 23. Conclusion}
\end{center}

\nopagebreak
Quantum field theory in the Standard Model requires gauge symmetry groups,
spontaneous symmetry breaking, and renormalization.} In 3D time these are not needed and
are unnecessary. The theory works much better without them. They are vestiges of an attempt
to build a unified field theory from the bottom up. The unification problem is so difficult,
however, that one must also proceed from the top down if one is to have any hope of succeeding.
This attacks the problem from both ends and recognizes the importance of having the
correct initial assumptions.  In place of these tenets of quantum field theory, 3D time
has extended particles, which produce cutoffs. Instead of juggling infinities,
these cutoffs eliminate the infinities before they appear.  Everything is finite and calculable, 
the way it should be, allowing the calculation of masses and coupling constants.

The Standard Model is just that --- a model that attempts to imitate a larger
and more comprehensive theory like 3D time. For example,
the Standard Model does not explain why there are three massive electroweak vectors
while one remains massless (the photon). There could be three massless vectors and  one massive;
or two massless vectors and two massive; or four massless vectors; or four massive ones.
3D time naturally produces one massless 
and three massive vectors because mass terms in the 6D curvature scalar have the form, e.g.
$A^\m \p_5^{\,2}A_\m$. This is zero for the photon, which is associated with the lowest 
order $l=m=0$ spherical harmonic that does not depend on the HD
coordinates $\w x^{\,5}$ or $\w x^{\,6}$. This singles out the photon for special treatment
and rightly so, as it is the fundamental in a series of harmonics. It has zero mass.
This mass term is not zero for all other spherical harmonics, which do depend on the 
higher-dimensional coordinates.

The reason 3D time can calculate masses is that we no longer have point particles. Instead, with its
extended particles come precisely calculated cutoffs of the order of the Planck mass.
The logarithm in the quantum mass correction equation converts these Planck masses 
to ordinary elementary particle masses, solving the hierarchy problem. 

If $SU(2)\times U(1)$ were correct, then $SU(5)$ or $SU(3)\times SU(3)$
would be right. But these theories have lead nowhere. In 3D time, the
$SU(2)\times U(1)$ structure of the standard model is replaced by 
spherical harmonics. Or more correctly, the spherical harmonics are
imitated by the $SU(2)\times U(1)$ structure of the standard model.

3D time is mathematically rigorous and precise. As a result it is highly organized, 
self-consistent, systematic, efficient and compact. This is unlike 
string theory, which seems to never end, and unlike the Standard Model, which must introduce
many of the terms in its Lagrangian ad hoc. These include the terms for Maxwell's equations
and the Dirac equation. These equations have their roots in experiment. The Standard Model
provides no explanation for why they should exist. It only assumes them. In contrast, 
3D time naturally produces these terms as part of the 6D curvature scalar.

The Standard Model is based on QED. QED is accepted because it works, explaining 
the gyromagnetic ratio of the electron. This despite the appearance of infinities.
3D time must be accepted because it also works, explaining 10 parameters of the Standard  Model.
It does this without the appearance of infinities. Not only this, but it actually eliminates
the infinities in QED!

One criticism of string theory is that it does not make predictions. String theorists claim it 
does make predictions. But how can we have faith in its predictions if it does not postdict 
or explain the existence of any phenomenon? It explains little about the known elementary 
particles. In contrast, 3D time has explained the existence of the Standard Model. 
Thus we can have faith in its predictions.

String theory makes the claim that it quantizes gravity. I claim that it does not even quantize
electromagnetism. QED, the standard for this, has infinities, which proves something is wrong.
By contrast, 3D time properly quantizes electroweak theory. It has no infinities. Everything is 
finite and calculable, the way it should be. As proof that it works it produces the masses and 
coupling constants of nine elementary particles. This is the example that should be set for
quantum gravity.

One does not need 10 or 11 dimensions to produce all the vectors needed for the Standard  Model.
If just one vector $\w g_{\mu 5}=A_\mu$ depends on the higher dimensional coordinates,
it will contain an infinite number of component vectors $A_\mu=\Sigma_{l,m}A_\mu^{lm}Y_{lm}$,
the first few of which, $A_\mu^{00},\ A_\mu^{1,-1},\ A_\mu^{10},\ A_\mu^{11}$, can be used
to describe the electroweak interaction. The remaining vectors with $l$ greater than one, are among the
theory's predictions. 

String theory must assume 10 or 11 dimensions, supersymmetry, strings and a large symmetry
group. Of these, 3D time assumes only 2 extra dimensions. That is all. 
And adding dimensions of time has worked before. Riemann tried to construct
a theory of gravitation using Riemannian geometry. This attempt failed 
because he used the three dimensions of space alone. It was only after Einstein added the
dimension of time that the theory worked. 

There is already more than enough evidence to imply 3D time is correct. 
As proof that 3D time is correct, I offer the fact that the solutions to its field equations are 
the known elementary particles. This means the theory explains why we see what we see for a 
vast body of observations. These include identifying the known elementary particles with specific
harmonics with well-defined quantum numbers; 
the absence of superpartners, which have never been detected; where the Uncertainty
Principle comes from; why there is minimal coupling
in the Standard Model; the existence of the Dirac equation; 
the apparent appearance of the symmetry group $SU(2)\times U(1)$; why the strong 
interactions are different than the electroweak interactions; why the photon is massless 
while the $W$ and $Z$ have mass; why the observed quarks and leptons exist;
where four-dimensional gravitation comes from (the 6D curvature scalar);
the existence of the $W$ and $Z$ along with the
terms that describe them; why there are three weak vectors and one electromagnetic vector;
interaction terms for vectors with spinors; why there appear to be three generations of leptons; 
why there are no four-point Fermi interactions; why there are no Feynman diagrams
with five or six external gluon lines; quark confinement; asymptotic freedom; chiral
symmetry breaking; how to make observations at the Planck distance; why there are two 
widely separated levels of mass in physics (the hierarchy problem);
the origin  of the Gell-Mann equation; how to construct an elementary particle;
whether the electron, muon, tau, the neutrinos, photon, $W$ and $Z$ have electric charge or not;
the mass of the electron; the mass of the photon; the value for the fine structure
constant; the masses of the muon and tau; the masses of the electron's, muon's and tau's 
neutrinos; why only the observed particle decays take place and not others; the masses of 
the $W$ and $Z$; and the masses of predicted particles allowing for a definite test of
the theory. This list of 45 phenomena explained by 3D time does not prove 3D time is correct.
It does prove, however, that it is far superior than anything else we have. The Standard Model
requires dozens of assumptions to explain the same phenomena. And
String theory does not explain any of these 45 items or much of anything else.
Despite a huge effort, String theory still seems to make no contact with reality.  
The bottom line is 3D time assumes much less than either the Standard Model 
or String theory and explains far more.

The short-distance behavior of the elementary particles is described by quantum mechanics. 
Therefore, one could say quantum mechanics is a property of the elementary particles. 
We have seen how general relativity produces the elementary particles. This means
general relativity produces quantum mechanics. 

It is well known that general relativity accounted for the expansion of the universe. I claim
it still does despite the current belief that the universe is expanding at an increasing rate.
This belief could easily be misplaced by an erroneous assumption. For example, everyone 
assumes Planck's constant has always had its current value. However, it
could easily have been much larger in the early universe. In geometric units, Planck's constant is equal to the
square of the Planck length, which is the radius of the HD sphere. If at some point in the early
universe all six dimensions were the same size, Planck's constant would have been much larger.
This would affect observations of the universe in the distant past, upon which the belief of
the universe expanding at an ever-increasing rate is based. In addition,
imaginary time appears to account for inflation:\,[67] the beginning of the universe may be represented 
by the north pole on a sphere with spherical coordinate $\theta$ for time and $\phi$
for space. As one moves away from this point the size of
the universe increases rapidly at first.

The theory of 3D time answers the question of why something exists rather than nothing.
Something will exist if and only if it is a solution to a field equation generated by the 6D
curvature scalar of 3D time.

General relativity is the most beautiful theory ever constructed.
With three dimensions of time, it also becomes the most practical,
accounting for the elementary particles as well as gravitation.
This is the best of all possible worlds.

Newton showed the force that pulls an apple from a tree is the same as that which 
keeps the earth in its orbit around the sun. We have now shown that
gravitation in 6D also accounts for the elementary particles and their interactions. 
It appears the entirety of the universe can be represented by a single letter:\,$\w R$.
Whether the solution to the field equations from the curvature scalar is a single electron
or the whole universe, it seems like everything in this world comes from general relativity.
The evidence suggests that the curvature scalar could be in control of this universe.

\clearpage
\begin{center}
{\bf References}
\end{center}

\begin{tabbing}
[1] T. Kaluza, Sitzsungsber.\ Preuss.\ 
     Akad.\ Wiss.\ Phys.\ Math.\ Kl.\ LIV (1921) 966. \\
{}[2] O. Klein, Z. Phys.\ 37 (1926) 895;
     Nature, 118 (1926) 516.\\
{}[3] S.L. Glashow, Nucl.\ Phys.\ 22 (1961) 579.\\
{}[4] S. Weinberg, Phys.\ Rev.\ Lett.\ 19 (1967) 1264.\\
{}[5] A. \=Salam, in: Elementary Particle Theory, ed.\ N. Svartholm
     (Almqvist, Forlag AB,\\ \>Stockholm, 1968) p.\ 367.\\
{}[6] G. 't Hooft, Nucl.\ Phys.\ B35 (1971) 167.\\
{}[7] A. Einstein, The Meaning of General Relativity 
       (Princeton Univ.\ Press, Princeton,\\
       \>1956) p.\ 129.\\
{}[8] J.A. Wheeler, Einstein's Vision (Springer, Berlin, 1968).\\
{}[9] A. Salam, Rev.\ Mod.\ Phys.\ 52 (1980) 525.\\
{}[10] J.M. Overduin and P.S. Wesson, Phys.\ Rep.\ 283 (1997) 303.\\
{}[11] B. De Witt, in: Relativity, Groups and Topology,
       eds. C. De Witt and B. De Witt\\
       \>(Gordon \& Breach, New York, 1964) p.\ 725.\\
{}[12] J. Rayski, Acta Phys.\ Pol.\ 27 (1965) 947; ibid 28 (1965) 87.\\
{}[13] R. Kerner, Ann.\ Inst.\ Henri Poincare 9 (1968) 143.\\
{}[14] A. Trautman, Rep.\ Math.\ Phys.\ 1 (1970) 29.\\
{}[15] Y.M. Cho, J.\ Math.\ Phys.\ 16 (1975) 2029.\\
{}[16] Y.M. Cho and P.G.O. Freund, Phys.\ Rev.\ D 12 (1975) 1711.\\
{}[17] P.S. Wesson, Gen.\ Rel.\ Grav.\ 22 (1990) 707.\\
{}[18] P.S. Wesson and J. Ponce de Leon, J. Math.\ Phys.\ 33 (1992) 3883.\\
{}[19] J. Ponce de Leon and P.S. Wesson, J. Math.\ Phys.\ 34 (1993) 4080.\\
{}[20] P.S. Wesson, J. Ponce de Leon, P. Lim and H. Liu, 
       Int.\ J.\ Mod.\ Phys.\ D 2 (1993) 163.\\
{}[21] P.S. Wesson, Mod.\ Phys.\ Lett.\ A 10 (1995) 15.\\
{}[22] S. Rippl, C. Romero and R. Tavakol, 
       Class.\ Quant.\ Grav.\ 12 (1995) 2411.\\
{}[23] P.S. Wesson et al., Int.\ J. Mod.\ Phys.\ A 11 (1996) 3247.\\
{}[24] B. Mashhoon, P.S. Wesson and H. Liu, 
       Univ.\ Missouri preprint (1996).\\
{}[25] T. Appelquist, A. Chodos and P.G.O. Freund eds., 
     Modern Kaluza-Klein Theories\\
     \>(Addison-Wesley, Menlo-Park, 1987).\\
{}[26] A. Salam and J. Strathdee, Ann.\ Phys.\ (NY) 141 (1982) 316.\\
{}[27] T. Appelquist and A. Chodos, Phys.\ Rev.\ Lett.\ 50 (1983) 141.\\
{}[28] T. Appelquist and A. Chodos, Phys.\ Rev.\ D 28 (1983) 772.\\
{}[29] L. Dolan and M.J. Duff, Phys. Rev.\ Lett.\ 52 (1984) 14.\\
{}[30] M.J. Duff, in:  Proc.\ 2nd Jerusalem Winter School of 
       Theoretical Physics, eds.\\
       \>T. Piran and S. Weinberg, (World Scientific, Singapore, 1986) p.\,40.\\
{}[31] M.J. Duff, B.E.W. Nilsson and C.N. Pope, 
       Phys.\ Rep.\ 130 (1986) 1.\\
{}[32] M.J. Duff, SISSA preprint hep-th/9410046 (1994).\\
{}[33] M.J. Duff, B.E.W. Nilsson, C.N. Pope and N.P. Warner, 
       Phys.\ Lett.\ B 149 (1984) 90.\\
{}[34] Y.M. Cho, Phys.\ Rev.\ D 35 (1987) 2628.\\
{}[35] Y.M. Cho, Phys.\ Lett.\ B 186 (1987) 38.\\
{}[36] Y.M. Cho and D.S. Kimm, J. Math.\ Phys.\ 30 (1989) 1570.\\
{}[37] Y.M. Cho, Phys.\ Rev.\ D 41 (1990) 2462.\\
{}[38] W. Thirring, Acta Phys.\ Austriaca, Suppl.\ 9 (1972) 256.\\
\end{tabbing}

\begin{tabbing}
{}[39] L.N.\= Chang, K.I. Macrae and F. Mansouri, 
       Phys.\ Rev.\ D 13 (1976) 235.\\
{}[40] G. Domokos and S. Kovesi-Domokos, 
       Il Nuovo Cimento A 44 (1978) 318.\\
{}[41] C.A. Orzalesi, Fortschr.\ Phys.\ 29 (1981) 413.\\
{}[42] C. Wetterich, Phys.\ Lett.\ B 113 (1982) 377.\\
{}[43] E. Cremmer and J. Scherk, Nucl.\ Phys.\ B 108 (1976) 409.\\
{}[44] E. Cremmer and J. Scherk, Nucl.\ Phys.\ B 118 (1977) 61.\\
{}[45] C.W. Misner, K.S. Thorne and J.A. Wheeler, Gravitation (W.H. Freeman and\\ 
     \>Company, San Francisco, 1973), Ch.\,41.\\
{}[46] W. Nahm, Nucl.\ Phys.\ B 135 (1978) 149.\\
{}[47] E. Witten, Nucl.\ Phys.\ B 186 (1981) 412.\\
{}[48] E. Cremmer, B. Julia and J. Scherk, Phys.\ Lett.\ B 76 (1978) 409.\\
{}[49] P.G.O. Freund and M.A. Rubin, Phys.\ Lett.\ B 97 (1980) 233.\\
{}[50] M.B. Green, J.H. Schwarz and E. Witten, Superstring theory
       (Cambridge Univ.\\ \> Press, Cambridge, 1987).\\
{}[51] P.D.B. Collins, A.D. Martin and E.J. Squires, Particle Physics and Cosmology\\
     \>(Wiley, New York, 1989).\\
{}[52] M.B. Green and J.H. Schwarz, Phys.\ Lett.\ B 149 (1984) 117.\\
{}[53] D.J. Gross, J.A. Harvey, E. Martinec and R. Rohm, 
         Phys.\ Rev.\ Lett.\ 54 (1985) 502.\\
{}[54] M.B. Green and J.H. Schwarz, Phys.\ Lett.\ B 109 (1982) 444.\\
{}[55] S.W. Hawking, A Brief History of Time (Bantam, 1988), p.\,134.\\
{}[56] G. 't Hooft, private communication.\\
{}[57] C.W. Misner, K.S. Thorne and J.A. Wheeler, Gravitation (W.H. Freeman and\\ \>Company, San Francisco, 1973), p.\,507.\\
{}[58] V.B. Berestetskii, E.M. Lifshitz and L.P. Pitaevskii, Relativistic
     Quantum Theory\\ \>Part~1 (Pergamon Press, New York, 1979), p.\,66.\\
{}[59] M.-Y. Han and Y. Nambu, Phys.\ Rev.\ 139B (1965) 1006.\\
{}[60] Y. Nambu and M.-Y. Han, Phys.\ Rev.\ D 10 (1974) 674.\\
{}[61] C. Itzykson and J. Zuber, Quantum Field Theory 
     (McGraw-Hill, New York, 1980),\\ \> Ch.\,7.\\
{}[62] J.D. Jackson, Classical Electrodynamics (John Wiley \& Sons, Inc\.,
     New York, 1975), \\ \>Chs.\,1, 3.\\
{}[63] S. Gasiorowicz, Quantum Physics (John Wiley \& Sons, Inc\., 
     New York, 1974), p.\,169.\\
{}[64] M.W. Grunewald, Phys.\ Rep.\ 322 (1999) 125.\\
{}[65] http://arxiv.org/abs/1407.6583\\
{}[66] http://arxiv.org/abs/1407.7494\\
{}[67] S.W. Hawking, A Brief History of Time (Bantam, 1988), p.\,138.\\
\end{tabbing}

\end{document}